\documentclass[11pt,letterpaper]{article}
\usepackage[T1]{fontenc}
\usepackage{amsmath}
\usepackage{amssymb}
\usepackage{amsthm}
\usepackage{graphicx}
\usepackage{booktabs}
\usepackage{array}
\usepackage{hyperref}
\usepackage{breakcites}
\usepackage{bbm}
\usepackage{xcolor}
\usepackage{subfig}
\usepackage{enumitem}
\usepackage{authblk}
\usepackage{natbib}
\usepackage{algorithm}
\usepackage{algpseudocode}

\newtheorem{theorem}{Theorem}[section]
\theoremstyle{remark}
\newtheorem{remark}[theorem]{Remark}
\theoremstyle{plain}
\newtheorem{corollary}[theorem]{Corollary}
\newtheorem{definition}[theorem]{Definition}
\newtheorem{proposition}[theorem]{Proposition}
\newtheorem{lemma}[theorem]{Lemma}

\newcommand{\dboxplus}{\mathop{\boxplus}\displaylimits}
\def\P{\mathbb{P}}
\newcommand{\E}{\mathbb{E}}
\newcommand{\Q}{\mathbb{Q}}
\newcommand{\R}{\mathbb{R}}
\newcommand{\Lset}{\mathbb{L}}
\newcommand{\eqd}{\overset{d}{=}}
\newcommand{\id}{\mathbbm{1}}
\newcommand{\bm}[1]{\boldsymbol{#1}}
\newcolumntype{L}[1]{>{\raggedright\arraybackslash}p{#1}}

\usepackage[margin=1.00in]{geometry}
\renewcommand{\d}{\, \mathrm{d}}

\title{Optimal robust comonotonic risk sharing under operational constraints}

\author[1,2]{Carole Bernard}
\author[3]{Christopher Blier-Wong}
\author[4]{Jean-Gabriel Lauzier}

\affil[1]{Grenoble École de Management, France}
\affil[2]{Vrije Universiteit Brussel, Belgium}
\affil[3]{Department of Statistical Sciences, University of Toronto, Canada}
\affil[4]{Department of Economics, Memorial University of Newfoundland, Canada}

\date{\today}

\begin{document}
\maketitle

\begin{abstract}
Risk-sharing arrangements such as reinsurance programs and catastrophe pools must satisfy operational constraints, including capacity limits, attachment bands, retention rules and cession limits. Under distortion risk measures, the unconstrained optimum assigns each loss layer to the participant for whom that layer is least costly, but operational constraints break this layerwise rule, and there is no general method to compute the constrained optimum.
We show that, for comonotonic allocations, the participants' distortion risk measures and the operational constraints are linear in the amount of each loss layer held by each participant, so the constrained risk-sharing problem on a discretized loss is a linear program that standard software can solve. Its dual gives supporting prices for the loss layers and the marginal value of relaxing each operational constraint. Where the set of binding constraints changes, the constrained optimal value can have a kink, and the smallest and largest optimal multipliers then give the one-sided effects of relaxing and tightening a constraint. To protect the allocation against disagreement between loss models, we also consider the robust version of the risk-sharing problem, that is, minimizing the worst-case total risk assessment over finitely many probability scenarios. This robust problem is again a linear program, and we quantify the cost of protection against model uncertainty from its optimal value. We illustrate the approach with a catastrophe reinsurance tower.
\end{abstract}


\section{Introduction}\label{sec:introduction}

In risk sharing, several participants divide a random loss so that each participant bears a manageable part. In catastrophe insurance, for instance, primary insurers, reinsurers, insurance-linked securities and public entities share the losses from hurricanes, earthquakes, floods and wildfires, because no single participant has the capital to bear these losses alone. In practice, participants cannot absorb arbitrary losses. An allocation must respect regulation, internal risk-management policies, and preferences or mandates specific to each participant. Capacity limits cap each participant's exposure at its available capital or at the amount it has committed by contract. Attachment requirements, retention rules, and cession limits determine which losses a participant may assume and how much risk may be transferred. This paper asks how such operational constraints can be built into optimal risk sharing, and how much each constraint costs.

Without operational constraints, optimal risk sharing under distortion risk measures has a simple solution. The optimization problem separates across layers of the aggregate loss, and the optimal allocation follows a layerwise rule: every layer goes to the participant for whom that layer is least costly \citep{embrechts2018quantile, lauzier2025risk}. Operational constraints break this separation. For instance, the cheapest participant may reach its capacity limit, and other participants must take the remaining layers. The allocation of one layer then depends on the allocation of every other layer. Other constraints link participants as well as layers: for example, a cession limit on the total amount ceded to several reinsurers restricts the layers held by all these reinsurers at once. Closed-form optimal allocations under such constraints are known only in special cases, such as the threshold rules of \citet{lo2017neyman} for constrained reinsurance. As far as we know, there is no general method in the literature to compute optimal allocations for many participants with distortion risk measures and general linear operational constraints. Yet the arrangements used in practice are constrained, so a numerical
method that solves the constrained problem is the missing piece for implementing optimal risk
sharing in practice.

We close this gap by showing that, on a finite loss grid, the constrained comonotonic risk-sharing problem is a linear program. We consider allocations in which each participant's loss is a nondecreasing function of the aggregate loss $S$ and equals zero when $S=0$, so that the allocation is comonotonic and satisfies the no-sabotage condition \citep{carlier2003insurance}. When $S$ is a continuous random variable, we discretize it on a finite grid ordered by loss magnitude. Consecutive grid points define the loss layers, and we describe an allocation by how much of each loss layer each participant holds. Because the allocation is comonotonic, each participant's distortion risk measure is a weighted sum of the amounts of the layers that the participant holds, with weights determined by the participant's distortion function and by the probability that the aggregate loss reaches each layer. The total risk assessment is therefore linear in the layer amounts. Capacity limits, attachment bands, retentions, quota shares and cession limits restrict how much of each layer each participant may hold, so these constraints are also linear in the layer amounts. Theorem~\ref{thm:lp} gives the resulting linear program for any number of participants, any distortion functions and any polyhedral system of operational constraints, so optimal allocations can be computed with standard linear-programming software even when no closed form is available.

Computing the optimal allocation is only one part of implementing risk sharing. Participants who accept part of the loss expect to be compensated, which requires a price for each loss layer. A participant deciding whether to raise capital, or members of a pool renegotiating a cession limit, need to know how much each operational constraint costs. Finally, catastrophe models from different vendors can assign quite different probabilities to the largest losses, so the allocation must perform well under more than one model. The linear-programming formulation provides all three. The dual of this linear program assigns a supporting price to every loss layer and a multiplier to every operational constraint. Theorem~\ref{thm:general-polyhedral-layer-pricing} shows how binding constraints move each layer price away from the cost of the cheapest participant on that layer. When the layer prices are normalized and nonincreasing, they define a probability pricing measure on the aggregate-loss grid (Sections~\ref{ss:dual_layer_prices} and~\ref{ss:constrained-price-system}). With the constraint multipliers, we can study the comparative statics of the constrained planner value, that is, the minimum total risk assessment under the operational constraints, which is the constrained counterpart of the infimal convolution. These comparative statics measure how much this value falls when a participant receives more capacity or a cession limit is relaxed, once the whole allocation is re-optimized. In economics, such questions are usually answered with the envelope theorem, whose classical form assumes that the planner value is differentiable in the constraint bounds. The constrained planner value is instead piecewise linear in these bounds. It can have kinks where the set of binding constraints changes, and at a kink the gain from relaxing a bound can differ from the loss from tightening it. Linear-programming sensitivity analysis gives both one-sided effects from the set of optimal multipliers.

To protect the allocation against disagreement between loss models, we choose a single allocation that minimizes the largest total risk assessment over a finite family of probability and distortion scenarios. Because the largest of finitely many linear functions can be minimized by linear programming, the robust problem is again a linear program, and the multipliers of the robust linear program keep their comparative-statics interpretation. Protection against model uncertainty has a price. The robust allocation is judged by its worst assessment across the candidate models, so this worst-case total risk assessment is at least the optimal value under each model taken alone. The robust layer prices are built only from the costs of the worst-case scenarios, so the layers to which these scenarios give more probability tend to become more expensive. Model uncertainty also changes which constraints bind and how much relaxing them is worth.

In Section~\ref{sec:numerical-examples}, we compute optimal allocations in a stylized catastrophe reinsurance tower with capacity limits, attachment bands, a retention corridor, a cession limit shared by two reinsurers, and a public backstop of limited size. The cession limit and the capacity limits link layers and participants, so the layerwise rule does not apply and the optimal allocation must be computed from the linear program. The capacity multipliers then give the marginal value of each participant's capacity. Following the marginal value of the backstop's capacity as that capacity grows gives a demand curve for public capacity, which drops sharply where the set of binding constraints changes, and which we compare under a single catastrophe model and in the worst case over three. To the best of our knowledge, this analysis is not available in the existing literature.

Optimal risk sharing originated in syndicate models of reinsurance \citep{borch1962equilibrium,wilson1968theory}. \citet{barrieu2005infconvolution} establish conditions for optimal transfers of risk to achieve the infimal convolution of monetary risk measures. The same problem has since been studied for other risk measures; see, e.g., \citet{acciaio2007optimal,xia2023mixed,ghossoub2024countermonotonic,lauzier2025risk}. For agents using Value-at-Risk and Expected Shortfall, \citet{embrechts2018quantile} characterize Pareto-optimal sharing, and \citet{embrechts2020quantilebased} extend the analysis to heterogeneous beliefs. Numerical methods for optimal risk sharing have largely considered the unconstrained case; for example, \citet{carlier2011numerical} formulate the problem with many agents and rank-linear utilities as a comonotonic variational problem and solve it by a conditional-gradient method. The optimal reinsurance literature also optimizes over classes of indemnity functions \citep{cai2008optimal,balbas2009optimal,cai2020optimal}. The approach closest to ours is that of \citet{zhuang2016marginal}, who use a function of the loss level taking values in $[0,1]$ specifying the marginal ceded fraction to parameterize the indemnity. Our layer amounts, divided by the layer sizes, are a discrete analogue of this marginal ceded fraction, which we use to share every layer among many participants under operational constraints rather than to design a single reinsurance contract.

Our formulation follows a familiar idea in risk optimization: Expected Shortfall and stochastic dominance constraints also become linear programs on finite distributions \citep{rockafellar2000optimization,rockafellar2002conditional,ogryczak2002dual,dentcheva2003optimization}, and dual representations of convex risk functions give optimality conditions for such programs \citep{ruszczynski2006optimization,shapiro2021lectures}. The ordered grid extends this idea to distortion risk measures under operational constraints. Our robust problem is a distributionally robust program with a finite ambiguity set \citep{bental2009robust,delage2010distributionally,wiesemann2014distributionally}; see \citet{kuhn2025dro} for a survey. \citet{asimit2019optimal} study optimal robust insurance contracts with a finite uncertainty set. \citet{bernard2025risk} also study risk sharing under distributional ambiguity, in a setting where the optimal allocation does not depend on the aggregate loss distribution; this property generally fails for our distortion risk measures. See also \citet{bernard2023robust}, \citet{cai2023distorted} and \citet{pesenti2025optimizing} for robust optimization of distortion risk measures.

The remainder of this paper is structured as follows. Section~\ref{sec:prelim} introduces the comonotonic allocation class, distortion risk measures, and finite-grid layer variables. In Section~\ref{sec:main}, we formulate the linear program, derive its supporting prices and study comparative statics. We treat the robust problem in Section~\ref{sec:robust-lp}. Section~\ref{sec:caps} specializes the constraints to capacity limits, under which the linear program is a transportation problem; there, we also give a greedy rule for the case of a single participant with a capacity limit. We present the case study in Section~\ref{sec:numerical-examples}, and Section~\ref{sec:discussion} concludes. Appendix~\ref{sec:discretization} shows that the grid values converge to the continuous planner value as the grid is refined, and Appendix~\ref{sec:proofs} collects linear-programming facts and the proofs.

\section{Preliminaries and notation}\label{sec:prelim}

\subsection{Admissible comonotonic allocations}\label{ss:admissible-comonotonic-allocations}

Let $(\Omega,\mathcal{F},\P)$ be an atomless probability space, and denote the bounded random variables by $\mathcal{X}:=L^\infty(\Omega,\mathcal{F},\P)$.  To impose the no-sabotage condition, every participant's allocation must be a nondecreasing function of the aggregate loss \mbox{\citep{carlier2003insurance}}; an assigned loss therefore cannot decline as the aggregate loss grows.  This condition covers standard arrangements such as retentions, quota shares, and deductibles on the aggregate loss. We thus seek optimal allocations that are comonotonic with the aggregate loss. 

\begin{definition}\label{def:comonotonicity}
Two random variables $X,Y\in \mathcal{X}$ are called comonotonic when
\begin{equation}\label{eq:comonotonicity}
\left(X(\omega)-X(\omega')\right)\left(Y(\omega)-Y(\omega')\right)\ge 0 \quad \text{for } (\P\otimes\P)\text{-a.s. } (\omega,\omega')\in\Omega\times\Omega.
\end{equation}
\end{definition}
Denneberg's Lemma \citep[Proposition~4.5]{denneberg1994nonadditive} states that $X$ and $Y$ are comonotonic precisely when some random variable $Z$ and nondecreasing functions $f_1,f_2:\R\to\R$ satisfy $X=f_1(Z)$ and $Y=f_2(Z)$ almost surely. A collection $(X_1,\dots,X_n)\in \mathcal{X}^n$ is comonotonic when all pairs $(X_i,X_j)$ are comonotonic. Equivalently, one common random variable $Z$ and nondecreasing functions $f_i:\R\to\R$ represent the allocation through $X_i=f_i(Z)$ for every $i\in[n]:=\{1,\dots,n\}$. In particular, one may take $Z=S=\sum_{i=1}^n X_i$.

Suppose agent $i\in[n]$ begins with a nonnegative loss $\zeta_i\in\mathcal X$. Let $S:=\sum_{i=1}^n\zeta_i$ be the aggregate loss and $\mathcal D_S:=[0,\mathop{\rm ess\,sup}S]$ its contract domain. The planner chooses a pure loss allocation, which distributes $S$ itself.  The class of admissible allocations is
\begin{equation}\label{eq:continuous-allocation-class}
\mathbb A_n(S)
:=
\left\{
\bm X\in\mathcal X^n:
\sum_{i=1}^n X_i=S\ \text{a.s.}
\right\}.
\end{equation}
The class of admissible comonotonic allocations is the subclass
\begin{equation}\label{eq:continuous-comonotonic-class}
\mathbb A_n^+(S) := \left\{\bm X\in\mathbb A_n(S): \begin{array}{l}
X_i=f_i(S),\quad f_i(0)=0,\quad i\in[n],\\
\displaystyle\sum_{i=1}^n f_i(s)=s,\quad s\in\mathcal D_S,\\
f_i\text{ is nondecreasing},\quad i\in[n]
\end{array}\right\}.
\end{equation}
Linear constraints on the allocations $\bm X$ are collected in an admissibility system $\Lset$. We then define
\begin{equation}\label{eq:continuous-constrained-comonotonic-class}
\mathbb A_n^{+,\Lset}(S) := \left\{\bm X\in\mathbb A_n^+(S):\bm X\text{ satisfies every constraint in }\Lset\right\}.
\end{equation}

An element of $\mathbb A_n(S)$ allocates the entire aggregate loss among the participants. The extra constraints in $\mathbb A_n^+(S)$ make every $X_i$ measurable with respect to $\sigma(S)$ and make $(X_1,\dots,X_n)$ comonotonic.  Monotonicity of each $f_i$ is the no-sabotage condition of \citet{carlier2003insurance}.  Moreover, $f_i(0)=0$ together with the monotonicity of $f_i$ ensures that every assigned loss $X_i$ is nonnegative.

\subsection{Distortion risk measures}
Throughout the paper, agents assess their risk through distortion risk measures. These measures originate in the dual theory of choice developed by \citet{yaari1987dual} and the premium principles of \citet{wang1996premium}. They are monotone, law-invariant, and comonotonic-additive functionals represented by Choquet integrals under distorted probabilities \citep{denneberg1994nonadditive}. Important regulatory measures, including Value-at-Risk and Expected Shortfall, belong to this class.

\begin{definition}\label{def:drm}
A distortion function is any nondecreasing map $h:[0,1]\to[0,1]$ satisfying $h(0)=0$ and $h(1)=1$. For a loss $X\in \mathcal{X}$, its distortion risk measure under $h$, denoted by $\rho_h$, is
\begin{equation}\label{eq:choquet}
\rho_h(X)  =\int X\d h\circ \P= \int_{0}^{\infty} h\left(\P(X>t)\right)\d t + \int_{-\infty}^{0}\left(h\left(\P(X>t)\right)-1\right)\d t.
\end{equation}
Here $\int X\d h\circ\P$ is the Choquet integral of $X$ with respect to $h\circ\P$, and both tail integrals are Lebesgue integrals with respect to $\d t$.
\end{definition}
We use the strict-tail convention $\P(X>t)$. The probability $\P(X>t)$ can differ from $\P(X\ge t)$ only at the (at most countably many) atoms of the law of $X$. Consequently, using weak tails instead would not change either integral in \eqref{eq:choquet}.

For $u\in(0,1)$, denote the lower $u$-quantile of $X$ by $F_X^{-1}(u):=\inf\{x\in\R:\P(X\le x)\ge u\}$. When $h$ is continuous, the distortion risk measure has the representation
\begin{equation}\label{eq:drm_quantile}
\rho_h(X)=\int_0^1 F_X^{-1}(u)\,\d\tilde h(u),
\end{equation}
where the dual distortion $\tilde h(u):=1-h(1-u)$ converts survival probabilities into lower quantiles. If $h$ is absolutely continuous, its dual has density $\gamma_h(u):=h'(1-u)$, so $\d\tilde h(u)=\gamma_h(u)\,\d u$. At an atom of the law of $X$, the quantile is constant across an interval of levels, and \eqref{eq:drm_quantile} always uses the fixed lower-quantile version $F_X^{-1}$.
Quantile representations for right- and left-continuous distortions are discussed by \citet{dhaene2012remarks}. A particularly relevant example is Expected Shortfall. Expected Shortfall at level $\alpha\in(0,1)$ has distortion $h$ and density $\gamma_h$ given by
\begin{equation}\label{eq:expected-shortfall}
h(x)=\min\left(\frac{x}{1-\alpha},\,1\right),
\qquad
\gamma_h(u)=\frac{1}{1-\alpha}\id_{\{u\in(\alpha,1)\}}.
\end{equation}

Four properties of distortion risk measures will be used repeatedly:
\begin{enumerate}
\item Monotonicity requires $\rho_h(X)\le \rho_h(Y)$ whenever $X\le Y$ a.s.
\item Cash additivity means $\rho_h(X+c)=\rho_h(X)+c$ for every $c\in\R$.
\item Law invariance gives $\rho_h(X)=\rho_h(Y)$ whenever $X\eqd Y$.
\item Comonotonic additivity gives $\rho_h(X+Y)=\rho_h(X)+\rho_h(Y)$ when $X$ and $Y$ are comonotonic.
\end{enumerate}

Concavity of $h$ makes $\rho_h$ coherent in the sense of \citet{artzner1999coherent} and consistent with convex order. These functionals are usually termed spectral risk measures; \citet{acerbi2002spectral} considers distortions continuous at zero. On general spaces, \citet{wang2020distortion} prove that concavity of $h$, convexity of $\rho_h$, and convex-order consistency are equivalent. Our finite-grid linear-programming results use only that $h$ is a distortion function and therefore also cover nonconcave (including convex) distortions, so that agents are not necessarily risk-averse.

The normalization $f_i(0)=0$ in \eqref{eq:continuous-comonotonic-class} separates the pure loss allocation from deterministic transfers. For constants $c_1,\dots,c_n$ with $\sum_{i=1}^n c_i=0$, cash additivity implies
$$\sum_{i=1}^n\rho_{h_i}(X_i+c_i)=\sum_{i=1}^n\rho_{h_i}(X_i).$$
Thus, zero-sum deterministic transfers may be chosen after the pure loss allocation has been solved. The pure loss allocation solves the planner's sharing problem, whereas selecting transfers or prices that implement it is a distinct task \citep{blier-wong2026transfers}.

\subsection{Planner's optimization}

The constrained planner value is the smallest total risk assessment that the participants can achieve by sharing $S$ under the constraints in $\Lset$; it is a constrained infimal convolution, defined as
\begin{equation}\label{eq:continuous-planner-value}
V(S;\Lset)
:=
\inf_{\bm X\in\mathbb A_n^{+,\Lset}(S)}
\sum_{i=1}^n\rho_{h_i}(X_i).
\end{equation}
For convex-order-consistent risk measures, the classical comonotonic improvement theorem replaces an allocation by a comonotonic one without increasing any agent's risk assessment \citep{landsberger1994comonotonic, ludkovski2008comonotonicity, filipovic2008optimal}. In a constrained problem, a sufficient condition for this improvement to preserve feasibility is that the nonempty feasible allocation set be componentwise convex-order solid \citep{blier-wong2026comonotonic}. When the improvement is admissible in the pure-loss class used here, a comonotonic optimum is also globally optimal. Otherwise, our formulation returns the best admissible comonotonic allocation without asserting global optimality over all allocations.

In the unconstrained comonotonic problem, define the infimal convolution by
$$\dboxplus_{i=1}^n\rho_{h_i}(S) := \inf_{\bm X\in\mathbb A_n^+(S)}\sum_{i=1}^n\rho_{h_i}(X_i).$$
For distortion risk measures, this infimal convolution equals the risk measure generated by the pointwise lower envelope of the distortions. For concave distortions, \citet[Lemma~2.2]{jouini2008optimal} establish the two-agent result, and \citet[Theorem~2]{ludkovski2009optimal} give an $n$-agent version, which yields the lower envelope when their transaction-cost coefficients vanish. Further formulations and extensions appear in \citet{boonen2015competitive,embrechts2018quantile,lauzier2025risk,ghossoub2026efficiency}. We state the version of \citet[Theorem~3]{lauzier2025risk}.
\begin{theorem}\label{thm:llw_th3}
Let continuous distortions $h_1,\dots,h_n$ induce the risk measures $\rho_{h_i}$, and let $S\in\mathcal X$ be a nonnegative aggregate loss on the atomless space $(\Omega,\mathcal F,\P)$. Then
$$\dboxplus_{i=1}^n \rho_{h_i}(S)=\rho_{h_\wedge}(S), \qquad h_\wedge(t):=\min_{1\le i\le n}h_i(t).$$
For each $x\in\mathcal D_S$, define the active set $M_x:=\{j\in[n]:h_j(\P(S>x))=h_\wedge(\P(S>x))\}$ of agents whose distortions attain $h_\wedge$ at $\P(S>x)$. The infimum is attained by an allocation $X_i=f_i(S)$, $i\in[n]$, with
$$f_i(x)=\int_0^x g_i(t)\,\d t, \qquad g_i(x)=\frac{1}{|M_x|}\id_{\{i\in M_x\}}, \qquad x\in\mathcal D_S.$$
\end{theorem}
The rule assigns each loss layer to the participant with the smallest distortion cost for that layer, that is, the one most willing to bear it.  In Theorem~\ref{thm:llw_th3}, the functions $f_i$ are nondecreasing, vanish at zero, and sum to the identity on $\mathcal D_S$; hence $(X_1, \dots, X_n)\in\mathbb A_n^+(S)$. In Section~\ref{sec:main}, we obtain the finite-state analogue of this lower-envelope identity.

\subsection{Finite-grid layer allocations}\label{ss:comonotonic-grid-constraints}

We next place the aggregate loss $S$ and the admissible comonotonic class $\mathbb A_n^+(S)$ on an ordered finite grid. Discretizing a continuous aggregate loss does not change the optimal value in the limit: for a bounded aggregate loss and grid-compatible constraints, the optimal value on the grid converges to the continuous planner value as the grid becomes finer (Appendix~\ref{sec:discretization}). For the remainder of this paper, we use the convention that capital letters denote cumulative or statewise quantities; lowercase letters denote increments and variables of the linear program.

Consider a grid of cumulative probabilities $0=P_0<P_1<\cdots<P_m=1$, derived below from the aggregate-loss distribution, and define the survival probabilities $\bar P_k:=1-P_{k-1}$ for $k=1,\dots,m$ and $\bar P_{m+1}:=0$. For a distortion $h$, set
\begin{equation}\label{eq:grid-distortion-coefficients}
H_k:=h(\bar P_k),\qquad H_{m+1}:=0,\qquad h_k:=H_k-H_{k+1},\qquad k=1,\dots,m.
\end{equation}
For absolutely continuous $h$, these differences satisfy
$$h_k=\int_{P_{k-1}}^{P_k}\gamma_h(u)\,\d u,\qquad k=1,\dots,m.$$
A fixed finite grid evaluates $h$ only at the survival probabilities $\bar P_k$ generated by the support of the aggregate loss.

Let $\Pi=\{A_1,\dots,A_m\}$ be a finite partition of $\Omega$, where $A_k\in\mathcal{F}$ and $\P(A_k)>0$, and put $\mathcal{F}_\Pi:=\sigma(\Pi)$. Assume that $S$ is $\mathcal F_\Pi$-measurable and equals $S_k$ on $A_k$, with $0=S_0\le S_1<\cdots<S_m.$ The value $S_0=0$ is only an auxiliary baseline (if zero belongs to the support of $S$, we can index that state as $S_1=0$; otherwise $S_1>0$). Equivalently,
\begin{equation}\label{eq:finite-state-aggregate-loss}
S=S^\Pi:=\sum_{k=1}^m S_k\,\id_{A_k}.
\end{equation}
We work with increments of these cumulative quantities. Define the loss increments and probabilities under the physical measure by
\begin{equation}\label{eq:finite-grid-loss}
s_k:=S_k-S_{k-1},\qquad p_k:=\P(A_k),\qquad P_k:=\sum_{\ell=1}^k p_\ell,
\end{equation}
for $k=1,\dots,m$, taking $P_0:=0$. The survival probabilities in \eqref{eq:grid-distortion-coefficients} are then $\bar P_k=1-P_{k-1}=\P(S\ge S_k)$. Thus $S_k$ is the cumulative aggregate loss at grid point $k$, and $s_k \ge 0$ is the size of layer $k$. Since the cell values $S_1<\cdots<S_m$ are distinct, $\sigma(S^\Pi)=\mathcal F_\Pi$. If $S$ is the sum of independent marginal losses, the convolution of their discretized laws produces the probability masses $(p_k)_{k=1}^m$.

Initialize $X_{0i}:=0$ for all $i\in[n]$. For $k=1,\dots,m$ and $i\in[n]$, let $X_{ki}$ be agent $i$'s cumulative allocation at grid point $k$. Its grid payoff is
\begin{equation}\label{eq:gridalloc}
X_i:=\sum_{k=1}^m X_{ki}\id_{A_k}.
\end{equation}
The amount of layer $k$ assigned to agent $i$ is the increment
$$x_{ki}:=X_{ki}-X_{k-1,i}.$$
Because the finite-grid version of \eqref{eq:continuous-planner-value} is restricted to allocations comonotonic with $S$, it imposes $x_{ki} \ge 0$. The cumulative amount given to agent $i$ through layer $k$ is therefore $X_{ki}=\sum_{\ell=1}^k x_{\ell i}$. Each layer must clear:
\begin{equation}\label{eq:layer-clearing}
\sum_{i=1}^n x_{ki}=s_k,\qquad k=1,\dots,m.
\end{equation}
Equation \eqref{eq:layer-clearing} gives $\sum_{i=1}^nX_{ki}=S_k$ at each grid point, so $\sum_{i=1}^nX_i=S$ almost surely. In addition, every sequence $k\mapsto X_{ki}$ is nondecreasing.

On this ordered grid, an individual distortion risk measure has the following linear expression in payoff increments.
\begin{lemma}\label{lem:comonotonic-grid-linearity}
Let $h$ be a distortion function and $Y=\sum_{k=1}^m Y_k\id_{A_k}$ a grid payoff satisfying $Y_1\le\cdots\le Y_m$. Choose any $Y_0\in\R$ and set $y_k:=Y_k-Y_{k-1}$, $k=1,\dots,m$. Then
\begin{equation}\label{eq:linearisedHg-layer}
\rho_h(Y)=Y_0+\sum_{k=1}^m y_k H_k=\sum_{k=1}^m Y_k h_k.
\end{equation}
\end{lemma}

\begin{proof}
Applying cash additivity and the Choquet representation \eqref{eq:choquet} to $Y-Y_1$, we get
$$\rho_h(Y)=Y_1+\sum_{k=2}^m (Y_k-Y_{k-1})H_k=Y_0+\sum_{k=1}^m y_kH_k.$$
The last equality follows from $y_1=Y_1-Y_0$ and $H_1=h(1)=1$. Because $H_{m+1}=0$, summation by parts then yields
$$Y_0+\sum_{k=1}^m y_kH_k=\sum_{k=1}^m Y_k(H_k-H_{k+1})=\sum_{k=1}^m Y_kh_k.$$
\end{proof}

Lemma~\ref{lem:comonotonic-grid-linearity} provides the evaluation formula for any admissible grid allocation. In \eqref{eq:linearisedHg-layer}, the layer form pairs increments $y_k$ with cumulative coefficients $H_k$, and the statewise form pairs cumulative values $Y_k$ with grid masses $h_k$. Cash additivity also shows that a deterministic transfer changes only $Y_0$, not the layer terms.

Fixing the grid turns the admissibility system $\Lset$ into a polyhedron $\Lset_\Pi$ described by finitely many linear rows. We state these constraints in the layer vector $\bm{x}$ and use the cumulative identities $X_{ki}=\sum_{\ell=1}^kx_{\ell i}$ where necessary. To construct a matrix representation, order the variables as
$$\bm{x}=(x_{11},\ldots,x_{m1},x_{12},\ldots,x_{m2},\ldots,x_{1n},\ldots,x_{mn})^\top$$
and let $\bm{e}_{ki}$ be the unit row that selects $x_{ki}$ for $k=1,\dots,m$ and $i=1,\dots,n$. Cumulative allocations then satisfy
$$X_{ki}=\sum_{\ell=1}^k \bm{e}_{\ell i}\bm{x}=\bm{E}_{ki}\bm{x},\qquad k=1,\dots,m,\quad i=1,\dots,n,$$
where $\bm{E}_{ki}:=\sum_{\ell=1}^k \bm{e}_{\ell i}=(0,\dots,0,1,\dots,1,0,\dots,0)$ has ones in the first $k$ positions of agent $i$'s block and zeros elsewhere. Choose matrices $A_{\Lset},E_{\Lset}$ and vectors $\bm{b}_{\Lset},\bm{r}_{\Lset}$ so that
$$\Lset_\Pi=\{\bm{x}\in\R^{mn}:E_{\Lset}\bm{x}=\bm{r}_{\Lset},\ A_{\Lset}\bm{x}\le \bm{b}_{\Lset}\}.$$
For brevity, write $A:=A_{\Lset}$ and $\bm b:=\bm b_{\Lset}$.
Appending the operational equality rows $E_{\Lset}\bm x=\bm r_{\Lset}$ to the layer-clearing equalities produces the full equality block
$$E^\top=\left(\sum_{i=1}^n \bm{e}_{1i}^\top,\ldots,\sum_{i=1}^n \bm{e}_{mi}^\top,E_{\Lset}^\top\right), \qquad \bm{r}^\top=(s_1,\ldots,s_m,\bm{r}_{\Lset}^\top).$$
Combining $\Lset_\Pi$ with layer clearing and monotonicity in aggregate loss gives the canonical feasible set
\begin{equation}\label{eq:grid-feasible-polyhedron}
\mathcal C_\Pi(\Lset)=\{\bm{x}\in\R_+^{mn}:E\bm{x}=\bm r,\, A\bm{x}\le\bm b\}.
\end{equation}
For the unconstrained problem, we have $\Lset_\Pi=\R^{mn}$.
Thus $E\bm{x}=\bm{r}$ contains both clearing and operational equalities, while $A\bm{x}\le\bm{b}$ contains the operational inequalities. Clearing and nonnegativity give $0\le x_{ki}\le s_k$ for all $k$ and $i$, so $\mathcal C_\Pi(\Lset)$, which is defined by finitely many linear equalities and non-strict inequalities, is a compact (possibly empty) polyhedron.

In Table~\ref{tab:constraint-catalogue}, we provide six of the most frequently used operational constraints for risk management applications. For the deductible row, agent $i$ can represent a reinsurer that pays nothing below the attachment level $a\in\mathcal D_S$. For the corridor row, $0\le\underline{c}\le\overline{c}\le1$ confines agent $i$'s cumulative allocation between $\underline{c}S_k$ and $\overline{c}S_k$. The joint cap bounds the total exposure of several agents, such as a cession limit shared by two reinsurers, and the layer corridor bounds the share of each layer held by agent $i$.

\begin{table}[!ht]
\centering
\setlength{\tabcolsep}{4pt}
\renewcommand{\arraystretch}{1.05}
\begin{tabular}{L{0.22\textwidth}L{0.22\textwidth}L{0.48\textwidth}}
\toprule
Constraint & Object & Matrix row or rows \\
\midrule
Terminal cap or layer capacity &
Endpoint exposure and total layer assignment &
$\sum_{\ell=1}^m\bm{e}_{\ell i}\bm{x}\le C_i$. \\
\addlinespace
Deductible or attachment &
Cumulative allocation $X_{ki}$ to reinsurer $i$ &
$\sum_{\ell=1}^k\bm{e}_{\ell i}\bm{x}=0$ for all $k=1,\dots,m$ with $S_k\le a$, entered as rows of $E\bm{x}=\bm{r}$. \\
\addlinespace
Corridor band &
Cumulative exposure $X_{ki}$ &
$-\sum_{\ell=1}^k\bm{e}_{\ell i}\bm{x}\le -\underline{c} S_k$ and $\sum_{\ell=1}^k\bm{e}_{\ell i}\bm{x}\le \overline{c} S_k$. \\
\addlinespace
Eligible layer or exhaustion band &
Layer amount $x_{ki}$ &
$x_{ki}=0$ for every pair of a layer and an agent outside the eligible band, entered as rows of $E\bm{x}=\bm r$. \\
\addlinespace
Joint cap &
Total exposure of a set $T$ of agents &
$\sum_{i\in T}\sum_{\ell=1}^m\bm{e}_{\ell i}\bm{x}\le C_T$. \\
\addlinespace
Layer corridor &
Layer amount $x_{ki}$ on a band of layers &
$\underline{c}\,s_k\le x_{ki}\le \overline{c}\,s_k$ for every layer $k$ in the band. \\
\bottomrule
\end{tabular}
\caption{Linear rows representing six operational constraints on finite-grid comonotonic allocations.}
\label{tab:constraint-catalogue}
\end{table}

Each entry of Table~\ref{tab:constraint-catalogue} becomes finitely many linear equalities or inequalities, imposing upper or lower bounds, that are appended directly to the matrices defining $\Lset_\Pi$. The formulation admits only constraints expressible through the layer variables on the aggregate-loss grid. A deductible on an initial loss $\zeta_i$, by contrast, depends on events in the finer state space on which $\zeta_i$ is measurable. Such a deductible is not admissible in our context, since the model permits only allocations that are measurable with respect to the aggregate loss $S$, not individual loss information. In Table~\ref{tab:notation}, we summarize the finite-grid notation.

\begin{table}[!ht]
\centering
\setlength{\tabcolsep}{4pt}
\renewcommand{\arraystretch}{1.08}
\begin{tabular}{L{0.18\textwidth}L{0.72\textwidth}}
\toprule
Symbol & Meaning \\
\midrule
$m$ & Number of grid layers. \\
$S_k$ & Cumulative aggregate loss at grid point $k=0,\dots,m$. \\
$s_k$ & Aggregate-loss increment $S_k-S_{k-1}$ on layer $k=1,\dots,m$. \\
$p_k$ & Physical probability mass of state $k=1,\dots,m$. \\
$P_k$ & Cumulative probability $\sum_{\ell=1}^k p_\ell$. \\
$\bar P_k$ & Survival probability $1-P_{k-1}$ before layer $k=1,\dots,m$. \\
$X_{ki}$ & Cumulative allocation $\sum_{\ell=1}^k x_{\ell i}$ to agent $i\in[n]$. \\
$x_{ki}$ & Allocation of layer $k=1,\dots,m$ assigned to agent $i\in[n]$. \\
$H_{ki}$ & Cumulative distortion tail cost $h_i(\bar P_k)$. \\
$h_{ki}$ & Difference $H_{ki}-H_{k+1,i}$. \\
\bottomrule
\end{tabular}
\caption{Notation for aggregate losses, allocations, and distortion costs on the finite grid.}
\label{tab:notation}
\end{table}

\section{Basic problem and supporting prices}\label{sec:main}

On the finite grid, the constrained comonotonic planner chooses a vector of layer amounts $\bm x$ from $\mathcal C_\Pi(\Lset)$. Once operational rows are imposed, we will show that their dual multipliers modify the cumulative distortion tail costs and the corresponding supporting prices. Throughout, we use standard linear-programming duality and sensitivity results \citep[Sections~4.3 and 5.2--5.3]{bertsimas1997introduction}. We recall these results in Appendix~\ref{sec:proofs}, where we also prove the results in Sections~\ref{sec:main}--\ref{sec:caps} that are not proved in the text.

\subsection{Constrained planner formulation}\label{ss:lp}

Consider agents $i\in[n]$ with distortion risk measures $\rho_{h_i}$. Applying \eqref{eq:grid-distortion-coefficients} to $h_i$ produces coefficients $H_{ki}$ and $h_{ki}$, indexed by layer $k$ and agent $i$. Lemma~\ref{lem:comonotonic-grid-linearity} shows that each risk measure is a fixed linear form in the layer variables: one additional unit of layer $k$ assigned to $i$ raises $\rho_{h_i}(X_i)$ by $H_{ki}$. Let $h_\wedge(u):=\min_{i\in[n]}h_i(u)$ be the lower-envelope distortion from Theorem~\ref{thm:llw_th3}. Applying \eqref{eq:grid-distortion-coefficients} to this envelope gives
$$H_k^\wedge:=\min_{i\in[n]}H_{ki},\qquad H_{m+1}^\wedge:=0,\qquad h_k^\wedge:=H_k^\wedge-H_{k+1}^\wedge,\qquad k=1,\dots,m.$$
Thus $h_\wedge$ is defined on $[0,1]$, while $H_k^\wedge$ and $h_k^\wedge$ are its cumulative and incremental coefficients on the grid.

Every admissible finite-grid comonotonic allocation satisfying $\Lset$ corresponds to a unique vector of nonnegative increments $\bm x\in\mathcal C_\Pi(\Lset)$. Conversely, any vector in $\mathcal C_\Pi(\Lset)$ uniquely determines a corresponding allocation. Lemma~\ref{lem:comonotonic-grid-linearity} further gives
$$\sum_{i=1}^n\rho_{h_i}(X_i)=\sum_{k=1}^m\sum_{i=1}^n H_{ki}x_{ki}.$$
The finite-grid planner value in \eqref{eq:continuous-planner-value} is therefore
\begin{equation}\label{eq:planner-grid}
V_\Pi(\Lset):=V(S^\Pi;\Lset)
= \min_{\bm{x}\in\mathcal C_\Pi(\Lset)}
\sum_{k=1}^m \sum_{i=1}^n H_{k i}\,x_{k i}.
\end{equation}
For a fixed admissibility system, we abbreviate $V_\Pi(\Lset)$ to $V_\Pi$. In Appendix~\ref{sec:discretization}, we show that these values converge to $V(S;\Lset)$ when the ordered grids are refined and the admissibility system is grid-compatible.

Let $\bm{c}$ denote the objective coefficient vector, ordered consistently with $\bm{x}$, with $c_{ki}=H_{ki}$. This gives the following finite-dimensional linear programming formulation.
\begin{theorem}\label{thm:lp}
Let $h_1,\dots,h_n$ be arbitrary distortion functions, and let the allocations in $\mathbb A_n^+(S)$ be represented on the finite grid by their layer vectors $\bm x$. If $\Lset_\Pi$ is a polyhedron given by finitely many linear equalities and inequalities, then \eqref{eq:planner-grid} is the finite-dimensional program
$$\min_{\bm{x}\in\R_+^{mn}} \bm{c}^\top\bm{x} \quad \textnormal{s.t.}\quad E\bm{x}=\bm{r}, \quad  A\bm{x}\le\bm{b}.$$
When $\mathcal C_\Pi(\Lset)$ is nonempty, program \eqref{eq:planner-grid} attains a finite value.
\end{theorem}

\begin{proof}
The objective of \eqref{eq:planner-grid} is linear in $\bm x$, and \eqref{eq:grid-feasible-polyhedron} gives $\mathcal C_\Pi(\Lset)=\{\bm{x}\in\R_+^{mn}: E\bm{x}=\bm r,\, A\bm{x}\le\bm b\}$. This set is compact, so the minimum is attained whenever it is nonempty.
\end{proof}

Because every admissible allocation is comonotonic, which fixes the ordering of each participant's payoffs, Theorem~\ref{thm:lp} requires neither concavity of the distortions nor convexity of the risk measures, and covers concave, nonconcave, and convex distortions alike.

\subsection{Layer decomposition and supporting prices}
\label{ss:dual_layer_prices}

To isolate the effect of operational constraints, first consider the unconstrained benchmark $\Lset_\Pi=\R^{mn}$. The only constraints are then clearing, $\sum_{i=1}^n x_{ki}=s_k$, and nonnegativity, $x_{ki}\ge0$. Program \eqref{eq:planner-grid} becomes
\begin{equation}\label{eq:layer_transport_lp}
V_\Pi = \min_{\bm{x}\ge\bm{0}}
\sum_{k=1}^m\sum_{i=1}^n H_{ki}x_{ki}
\quad\textnormal{s.t.}\quad
\sum_{i=1}^n x_{ki}=s_k, \quad k=1,\dots,m.
\end{equation}

\begin{corollary}
\label{cor:layer_assignment}
Suppose $\Lset_\Pi=\R^{mn}$. An optimal solution $\bm{x}$ may be chosen with indices $i^\star(k)\in\arg\min_{i\in[n]} H_{ki}$ such that $x_{k,i^\star(k)}=s_k$ and $x_{ki}=0$ whenever $i\neq i^\star(k)$. Its value is
\begin{equation}\label{eq:value_closed_form}
V_\Pi = \sum_{k=1}^m s_k H_k^\wedge.
\end{equation}
\end{corollary}

The minimizer $i^\star(k)$ may not be unique. If several agents attain $H_k^\wedge$, the planner can give the whole layer to any one of them or divide it equally among them; either choice attains the same layer cost. Corollary~\ref{cor:layer_assignment} thus recovers on the grid the lower-envelope allocation of Theorem~\ref{thm:llw_th3}, due to \citet{lauzier2025risk}.

The dual of this program gives the prices of the loss layers that support the unconstrained allocation, and these prices serve as the benchmark for the constrained prices of Section~\ref{ss:constrained-price-system}. To form the dual of \eqref{eq:layer_transport_lp}, attach an unrestricted variable $\lambda_k\in\R$ to each clearing equation $\sum_{i=1}^n x_{ki}=s_k$. Consider the candidate prices
$$\lambda_k^\star:=H_k^\wedge,\qquad k=1,\dots,m,\qquad \lambda_{m+1}^\star:=0.$$
The next corollary identifies the dual and verifies optimality of this sequence.

\begin{corollary} \label{cor:dual_layer_prices}
The dual of the minimization program in
Corollary~\ref{cor:layer_assignment} is
\begin{equation}
\max_{\lambda\in\R^m} \sum_{k=1}^m \lambda_k s_k \qquad \textnormal{s.t.} \qquad \lambda_k\le H_{ki}, \quad k=1,\ldots,m,\ i\in[n].
\end{equation}
Its optimal value is $V_\Pi$, and $(\lambda_k^\star)_{k=1}^m$
defines an optimal dual solution. Further, we have that
$$\Q^\star(A_k):=q_k^\star = \lambda_k^\star-\lambda_{k+1}^\star = h_k^\wedge, \qquad k=1,\ldots,m$$
defines a probability measure on $(\Omega,\mathcal F_\Pi)$.
\end{corollary}

We call the probability measure $\Q^\star$ the unconstrained pricing measure. For an $\mathcal F_\Pi$-measurable claim $Y=\sum_{k=1}^m Y_k\id_{A_k}$, set $Y_0:=0$. Then
$$\E^{\Q^\star}[Y]=\sum_{k=1}^m q_k^\star Y_k =\sum_{k=1}^m\lambda_k^\star(Y_k-Y_{k-1}).$$
This identity gives an Arrow--Debreu state-price representation on the finite grid \citep{arrow1964role,debreu1959theory}. The mass $q_k^\star$ is the normalized price of state $A_k$, and valuation by these state prices coincides with valuation by the layer prices $\lambda_k^\star$. \citet[Section~5]{embrechts2018quantile} use this interpretation, under additional assumptions, to construct an equilibrium pricing rule in quantile-based risk sharing. Since $\lambda_k^\star$ is dual to the clearing right-hand side $s_k$, it measures the marginal cost of one extra unit of loss in layer $k$.

In the unconstrained layer program \eqref{eq:layer_transport_lp}, $\bm{\lambda}^\star$ is the lower envelope $H_k^\wedge$ of cumulative distortion tail costs. These prices generate the grid measure $\Q^\star$, under which \eqref{eq:value_closed_form} prices the aggregate loss at $V_\Pi$. For concave distortions, $\Q^\star$ is the finite-grid counterpart of the lower-envelope pricing measure derived in \citet[Appendix~B]{blier-wong2026transfers}. If endowments are $\mathcal F_\Pi$-measurable, $\Q^\star$ also prices each agent's change in position: if $\E^{\Q^\star}[\zeta_i-X_i]$ is positive, agent $i$ pays this amount for loss relief, and if it is negative, agent $i$ receives its absolute value for taking additional loss.

\subsection{Constrained price systems}\label{ss:constrained-price-system}

Recall that $\Lset_\Pi$ consists of operational equalities $E_{\Lset}\bm{x}=\bm{r}_{\Lset}$ and inequalities $A\bm{x}\le\bm{b}$. Here $A\in\R^{r\times mn}$ and $\bm{b}\in\R^r$. We keep the notation $E_{\Lset}\bm{x}=\bm{r}_{\Lset}$, where $E_{\Lset}\in\R^{e_{\Lset}\times mn}$ and $\bm{r}_{\Lset}\in\R^{e_{\Lset}}$, to separate operational equalities from the full block $E\bm{x}=\bm{r}$ that also contains clearing. A row may involve a single variable $x_{ki}$, every layer assigned to one agent, or variables from several agents and layers. The constrained problem is
$$\begin{aligned}
(P_\Pi(\bm{b},\bm{r}_{\Lset}))\qquad
V_\Pi(\bm{b},\bm{r}_{\Lset})
:=\min_{\bm{x}\in\R_+^{mn}}\quad& \sum_{k=1}^m\sum_{i=1}^n H_{ki}x_{ki}\\
\textnormal{s.t.}\quad& \sum_{i=1}^n x_{ki}=s_k,\qquad k=1,\dots,m,\\
&E_{\Lset}\bm{x}=\bm{r}_{\Lset},\\
&A\bm{x}\le \bm{b}.
\end{aligned}$$
With operational equality levels held fixed, we abbreviate the primal and its value to $P_\Pi(\bm{b})$ and $V_\Pi(\bm{b})$; the dual below is likewise denoted $D_\Pi(\bm{b})$.

Let $\bm\lambda$ and $\bm\eta$ be the unrestricted multipliers on clearing and the operational equalities, respectively, and let $\bm\mu\in\R_+^r$ multiply the operational inequalities. The Lagrangian of this primal is
\begin{equation}\label{eq:constrained-primal-lagrangian}
\begin{aligned}
\mathcal L_{P_\Pi}(\bm x;\bm\lambda,\bm\eta,\bm\mu)
:={}&\sum_{k=1}^m\sum_{i=1}^nH_{ki}x_{ki}
+\sum_{k=1}^m\lambda_k\left(s_k-\sum_{i=1}^n x_{ki}\right)\\
&+\bm\eta^\top(\bm r_{\Lset}-E_{\Lset}\bm x)
+\bm\mu^\top(A\bm x-\bm b).
\end{aligned}
\end{equation}
The coefficient of $x_{ki}$ in the Lagrangian is $-\lambda_k$ plus the shadow-adjusted cost
$$H_{ki}-(E_{\Lset}^\top\bm{\eta})_{ki} +(A^\top\bm{\mu})_{ki}, \qquad k=1,\dots,m,\quad i\in[n].$$
Let $e_{j,ki}$ denote the coefficient of $x_{ki}$ in operational equality $j$, and let $a_{\ell,ki}$ denote its coefficient in operational inequality $\ell$, so that $(E_{\Lset}^\top\bm{\eta})_{ki}=\sum_{j=1}^{e_{\Lset}}e_{j,ki}\eta_j$ and $(A^\top\bm{\mu})_{ki}=\sum_{\ell=1}^r a_{\ell,ki}\mu_\ell$.

\begin{proposition}
\label{prop:general-polyhedral-layer-dual}
The program dual to $P_\Pi(\bm{b},\bm{r}_{\Lset})$ is
$$\begin{aligned}
(D_\Pi(\bm{b},\bm{r}_{\Lset}))\qquad
\max_{\substack{\bm{\lambda}\in\R^m,\ \bm{\eta}\in\R^{e_{\Lset}},\\
\bm{\mu}\in\R_+^r}}\quad&
\sum_{k=1}^m \lambda_k s_k+\bm{r}_{\Lset}^\top\bm{\eta}-\bm{b}^\top\bm{\mu}\\
\textnormal{s.t.}\quad&
\lambda_k\le H_{ki}-(E_{\Lset}^\top\bm{\eta})_{ki}+(A^\top\bm{\mu})_{ki},
\qquad k=1,\dots,m,\ i\in[n].
\end{aligned}$$
When $\mathcal C_\Pi(\Lset)$ is nonempty, strong duality holds.
\end{proposition}

An operational constraint may involve the amounts of several layers or several agents, so the amount assigned on one layer can restrict the amounts that are feasible on other layers. The next theorem characterizes the resulting supporting prices at a primal--dual optimum.

\begin{theorem}
\label{thm:general-polyhedral-layer-pricing}
Suppose $P_\Pi(\bm{b},\bm{r}_{\Lset})$ is feasible, and choose any primal--dual optimum $(\bm{x}^{\Lset},\bm{\lambda}^{\Lset},\bm{\mu}^{\Lset},\bm{\eta}^{\Lset})$. For each layer $k=1,\dots,m$ with $s_k>0$ we have
$$\lambda_k^{\Lset} =\min_{i\in[n]} \left( H_{ki}-(E_{\Lset}^\top\bm{\eta}^{\Lset})_{ki} +(A^\top\bm{\mu}^{\Lset})_{ki} \right),$$
and
$$\lambda_k^{\Lset} =H_{ki}-(E_{\Lset}^\top\bm{\eta}^{\Lset})_{ki} +(A^\top\bm{\mu}^{\Lset})_{ki}$$
for every $(k,i)\in\{1,\dots,m\}\times[n]$ with $x_{ki}^{\Lset}>0$.
\end{theorem}

Theorem~\ref{thm:llw_th3} determines the unconstrained allocation from the lower envelope of the primitive cumulative distortion costs $H_{ki}$. The present theorem preserves this lower-envelope structure under operational constraints. Inequality row $\ell$ shifts agent $i$'s cost of layer $k$ by $a_{\ell,ki}\mu_\ell^{\Lset}$, while equality row $j$ shifts it by $-e_{j,ki}\eta_j^{\Lset}$. A positive coefficient $a_{\ell,ki}$ lifts the corresponding curve of cumulative tail costs when $\mu_\ell^{\Lset}>0$, whereas a negative coefficient shifts it downward. Equality constraints can shift a curve in either direction because $\eta_j^{\Lset}$ can have either sign. The price $\lambda_k^{\Lset}$ is still the lower envelope of these adjusted costs. Complementary slackness assigns positive layer mass only to agents on that envelope, who price the adjusted layer most cheaply. A layer can therefore be split only among agents whose adjusted costs tie at the envelope. Solving the primal--dual linear program therefore both finds the optimal allocation and computes the multipliers that transform the primitive envelope into its constrained counterpart.

We now ask when supporting layer prices generate a pricing measure $\Q$ on $\Pi$, so that every $\mathcal F_\Pi$-measurable contract $Y$ is priced by $\E^\Q[Y]$. Given cumulative prices $\bm{\lambda}\in\R^m$, append $\lambda_{m+1}:=0$ and define event masses by
\begin{equation}\label{eq:pricing-mass-convention}
q_k(\bm{\lambda}):=\lambda_k-\lambda_{k+1},\qquad k=1,\dots,m.
\end{equation}
If $\lambda_1=1$, then $\bm{\lambda}$ is said to be normalized. If $\lambda_1\ge\cdots\ge\lambda_m\ge\lambda_{m+1}$, where $\lambda_{m+1}=0$ as above, then $\bm{\lambda}$ is said to be nonincreasing. When both properties hold, the masses $q_k(\bm{\lambda})$ are nonnegative and sum to one. They therefore define the probability measure $\Q_{\lambda}$ through $\Q_{\lambda}(A_k)=q_k(\bm{\lambda})$.

For a grid claim $Y=\sum_{k=1}^mY_k\id_{A_k}$, take $Y_0:=0$ and $y_k:=Y_k-Y_{k-1}$. The pricing functional for the layers associated with $\bm{\lambda}$ is
$$\mathsf{P}_{\bm{\lambda}}(Y) :=\sum_{k=1}^m\lambda_k y_k =\sum_{k=1}^m q_k(\bm{\lambda})Y_k.$$
If $\bm{\lambda}$ is normalized and nonincreasing, this functional has the representation
$$\mathsf{P}_{\bm{\lambda}}(Y)=\E^{\Q_{\lambda}}[Y].$$
Theorem~\ref{thm:general-polyhedral-layer-pricing} always makes $\bm{\lambda}^{\Lset}$ a vector of supporting marginal layer prices for $P_\Pi(\bm{b},\bm{r}_{\Lset})$, with masses $q_k^{\Lset}:=q_k(\bm{\lambda}^{\Lset})$. Without constraints, these masses are those of the probability pricing measure $\Q^\star$ in Corollary~\ref{cor:dual_layer_prices}. If the constraints shift the costs of only some pairs of layers and agents, the adjusted envelope, and hence the layer prices, may no longer be nonincreasing. Capacity constraints are important exceptions (Corollary~\ref{cor:scarcity-adjusted-pricing}).

For multipliers $\bm{\mu}\ge\bm{0}$ and $\bm{\eta}\in\R^{e_{\Lset}}$, define the adjusted lower envelope by
$$\lambda_k^{\bm{\mu},\bm{\eta}} := \min_{i\in[n]} \left( H_{ki}-(E_{\Lset}^\top\bm{\eta})_{ki} +(A^\top\bm{\mu})_{ki} \right), \qquad k=1,\dots,m,$$
and set $\lambda_{m+1}^{\bm{\mu},\bm{\eta}}:=0$. Summation by parts with the masses $q_k(\bm{\lambda}^{\bm{\mu},\bm{\eta}})$ prices the aggregate loss as
\begin{equation}\label{eq:shadow-adjusted-layer-price}
\mathsf{P}_{\bm{\lambda}^{\bm{\mu},\bm{\eta}}}(S)
:=\sum_{k=1}^m \lambda_k^{\bm{\mu},\bm{\eta}}s_k
=\sum_{k=1}^m q_k(\bm{\lambda}^{\bm{\mu},\bm{\eta}}) S_k .
\end{equation}
Substituting the adjusted envelope into the dual gives
\begin{equation}\label{eq:constrained-value-shadow-price}
V_\Pi(\bm{b},\bm{r}_{\Lset})
=
\sup_{\substack{\bm{\mu}\ge\bm{0},\\
\bm{\eta}\in\R^{e_{\Lset}}}}
\left(
\mathsf{P}_{\bm{\lambda}^{\bm{\mu},\bm{\eta}}}(S)
+\bm{r}_{\Lset}^\top\bm{\eta}
-\bm{b}^\top\bm{\mu}
\right).
\end{equation}
Equation~\eqref{eq:constrained-value-shadow-price} separates two parts of the constrained value. The adjusted envelope gives the price of the aggregate loss under the supporting layer prices. The last two terms account for the prescribed equality levels $\bm r_{\Lset}$ and inequality bounds $\bm b$. By Proposition~\ref{prop:general-polyhedral-layer-dual}, every optimal pair $(\bm{\mu}^{\Lset},\bm{\eta}^{\Lset})$ attains the supremum. At such a pair, the last two terms are exactly the difference between the planner value and the value assigned to $S$ by the supporting layer prices. In the unconstrained problem, the price of $S$ under the supporting layer prices equals the planner value (Corollary~\ref{cor:dual_layer_prices}); with operational constraints, the two can differ.

\subsection{Comparative statics of operational constraints}\label{ss:operational-sensitivity}

We now study how much the minimum total risk assessment falls when a participant receives more capacity or a cession limit is relaxed, after the whole allocation is re-optimized. These are comparative statics of the constrained planner value $V_\Pi$. In economics, such questions are usually answered with the envelope theorem, whose classical form assumes that the planner value $V_\Pi$ is differentiable.
Here $V_\Pi$ is convex and piecewise linear in the bounds $(\bm b,\bm r_{\Lset})$ and can have kinks where the set of binding constraints changes, so the classical envelope theorem cannot be applied at those points. We use linear-programming duality and sensitivity analysis instead. The next result identifies the binding constraints and gives the marginal effects where $V_\Pi$ is differentiable.

\begin{corollary}\label{cor:operational-shadow-values}
Suppose $P_\Pi(\bm{b},\bm{r}_{\Lset})$ is feasible, and let $(\bm{x}^{\Lset},\bm{\lambda}^{\Lset},\bm{\mu}^{\Lset},\bm{\eta}^{\Lset})$ be primal--dual optimal. Then $\mu_\ell^{\Lset}>0$ implies $(A\bm{x}^{\Lset})_\ell=b_\ell$ for every $\ell\in\{1,\dots,r\}$. Moreover, $V_\Pi$ is convex in $(\bm{b},\bm{r}_{\Lset})$ and nonincreasing in $\bm{b}$, with
$$\bigl(-\bm{\mu}^{\Lset},\bm{\eta}^{\Lset}\bigr) \in\partial V_\Pi(\bm{b},\bm{r}_{\Lset}).$$
If $V_\Pi$ is differentiable at $(\bm{b},\bm{r}_{\Lset})$ then
$$\frac{\partial V_\Pi}{\partial b_\ell} =-\mu_\ell^{\Lset}, \qquad \frac{\partial V_\Pi}{\partial r_{\Lset,j}} =\eta_j^{\Lset},$$
for $\ell=1,\dots,r$ and $j=1,\dots,e_{\Lset}$.
\end{corollary}

Corollary~\ref{cor:operational-shadow-values} gives derivatives only where $V_\Pi$ is differentiable. Since $V_\Pi$ is polyhedral, it can have kinks where the set of binding constraints changes, and at a kink, relaxing a bound and tightening it can change the value at different rates. The next result gives both one-sided derivatives along a direction $\bm d\in\R^r$ of change in the inequality bounds. Let $\operatorname{dom}V_\Pi$ denote the set of bounds $\bm b$ for which $P_\Pi(\bm b)$ is feasible, and hence has a finite value by Theorem~\ref{thm:lp}. For $\bm b\in\operatorname{dom}V_\Pi$, let $\mathcal M(\bm b)$ be the set of inequality multipliers $\bm\mu$ that belong to an optimal solution of $D_\Pi(\bm b)$.

\begin{proposition}\label{prop:one-sided-sensitivity}
Let $\bm b\in\operatorname{dom}V_\Pi$ and $\bm d\in\R^r$ be such that $\bm b+\varepsilon\bm d\in\operatorname{dom}V_\Pi$ for all sufficiently small $\varepsilon>0$. Then
$$\lim_{\varepsilon\downarrow0}\frac{V_\Pi(\bm b+\varepsilon\bm d)-V_\Pi(\bm b)}{\varepsilon}=-\min_{\bm\mu\in\mathcal M(\bm b)}\bm d^\top\bm\mu.$$
If also $\bm b-\varepsilon\bm d\in\operatorname{dom}V_\Pi$ for all sufficiently small $\varepsilon>0$, then
$$\lim_{\varepsilon\downarrow0}\frac{V_\Pi(\bm b)-V_\Pi(\bm b-\varepsilon\bm d)}{\varepsilon}=-\max_{\bm\mu\in\mathcal M(\bm b)}\bm d^\top\bm\mu,$$
so $\varepsilon\mapsto V_\Pi(\bm b+\varepsilon\bm d)$ is differentiable at zero if and only if $\bm d^\top\bm\mu$ is the same for every $\bm\mu\in\mathcal M(\bm b)$.
\end{proposition}

At a kink, an arbitrary optimal multiplier therefore need not measure the gain from relaxing a bound. Taking $\bm d$ as the $\ell$th unit vector in Proposition~\ref{prop:one-sided-sensitivity}, the gain from relaxing bound $\ell$ is the smallest optimal multiplier $\min_{\bm\mu\in\mathcal M(\bm b)}\mu_\ell$, and the loss from tightening it is the largest, $\max_{\bm\mu\in\mathcal M(\bm b)}\mu_\ell$. Section~\ref{sec:caps} specializes these results to capacity limits.

\section{Robust problem}\label{sec:robust-lp}

Section~\ref{sec:main} treats the grid probabilities $p_k=\P(A_k)$ as known. In practice, insurers often have several candidate probability models, estimated from historical observations or produced by catastrophe models. Tail data are sparse, so these models can assign very different probabilities to the largest losses, and a treaty optimized under one model may perform badly under another. \citet{embrechts2022robustness} show that small model errors can substantially change the risk of a position optimized under a reference model. We therefore let the planner choose one allocation for all candidate models and minimize the largest of the resulting total risk assessments. This is a distributionally robust program with a finite ambiguity set (see Section~\ref{sec:introduction} for related work). We keep the ambiguity set finite because the problem then remains a linear program. Its dual assigns a weight to each candidate model, and only models that attain the worst case at the robust optimum can receive positive weight.

Protection against model uncertainty has a price. The robust allocation is judged by its worst assessment across the candidate models, so the robust value is at least the optimal value under each candidate model taken alone. Section~\ref{subsec:robust-dual-weights} shows that the robust layer prices are built only from the costs of the scenarios that attain the worst case, so the layers to which these scenarios give more probability tend to become more expensive. Model uncertainty also changes which constraints bind and how much relaxing them is worth, and we can use the multipliers of the robust linear program to measure these effects as in Section~\ref{ss:operational-sensitivity}.

\subsection{Scenarios and robust program}\label{subsec:robust-formulation}

Suppose we have $M\ge1$ scenarios, indexed by $a\in[M]:=\{1,\dots,M\}$. Scenario $a$ specifies a probability law $\P^a$ and a distortion function $h_i^a$ for each agent $i\in[n]$; we write $\theta^a:=(\P^a,h_1^a,\dots,h_n^a)$ and $\Theta_M:=\{\theta^1,\dots,\theta^M\}$. All scenarios use the same grid, operational constraints and allocation variables. The partition can combine the support points of all scenarios, so a cell may have zero probability under some scenarios. Scenario $a$ enters the robust problem only through the cumulative tail costs
$$H_{ki}^a:=h_i^a(\bar P_k^a),\qquad \bar P_k^a:=\P^a(S\ge S_k),\qquad k=1,\dots,m,\quad i\in[n],$$
and we set $H_{m+1,i}^a:=0$. Since every $h_i^a$ is a distortion function (Definition~\ref{def:drm}), we have that $H_{1i}^a=1$ and $k\mapsto H_{ki}^a$ is nonincreasing. The scenarios can thus differ in their probability laws, in the agents' distortions, or in both. If only the laws differ, we have $H_{ki}^a=h_i(\bar P_k^a)$. Agents may also use different laws within a scenario, as with heterogeneous beliefs. Agent $i$'s costs are then $H_{ki}^a=h_i^a\bigl(\P_i^a(S\ge S_k)\bigr)$, and the results below apply without change, including for $M=1$. Distortion ambiguity sets built from shape constraints and elicited preferences, as in \citet{wangxu2023preference}, fit in this framework once they are represented by finitely many distortions.

The planner chooses one allocation $\bm x\in\mathcal C_\Pi(\Lset)$ and evaluates it under every scenario. Let $\rho_i^a$ be the distortion risk measure defined by $h_i^a$ under $\P^a$. All allocations in $\mathcal C_\Pi(\Lset)$ are comonotonic, so Lemma~\ref{lem:comonotonic-grid-linearity} applies under each scenario and gives
$$\sum_{i=1}^n\rho_i^a(X_i)=\sum_{k=1}^m\sum_{i=1}^nH_{ki}^a x_{ki},\qquad a\in[M].$$
The robust value is the smallest worst-case total risk assessment,
$$V_{\Pi}(\bm{b},\bm{r}_{\Lset};\Theta_M):=\min_{\bm{x}\in\mathcal C_\Pi(\Lset)}\ \max_{a\in[M]}\ \sum_{k=1}^m\sum_{i=1}^n H_{ki}^a x_{ki}.$$
Adding one variable $\tau$ and one inequality per scenario turns this min--max problem into a linear program; this is the standard epigraph reformulation.

\begin{theorem}
\label{thm:robust-lp}
The robust value $V_{\Pi}(\bm{b},\bm{r}_{\Lset};\Theta_M)$ equals the value of the linear program
\begin{equation}\label{eq:robust-solver-primal}
\begin{aligned}
(P_{\Pi}(\bm{b},\bm{r}_{\Lset};\Theta_M))\qquad
\min_{\bm{x}\in\R_+^{mn},\ \tau\in\R}\quad
& \tau\\
\textnormal{s.t.}\quad
& \sum_{i=1}^n x_{ki}=s_k,\qquad k=1,\dots,m,\\
& E_{\Lset}\bm{x}=\bm{r}_{\Lset},\\
& A\bm{x}\le\bm{b},\\
& \sum_{k=1}^m\sum_{i=1}^n H_{ki}^a x_{ki}\le \tau,
\qquad a=1,\dots,M.
\end{aligned}
\end{equation}
When $\mathcal C_\Pi(\Lset)$ is nonempty, program \eqref{eq:robust-solver-primal} attains a finite value.
\end{theorem}

Robustness adds only one variable and $M$ rows to the program~\eqref{eq:planner-grid}; clearing, the operational constraints and nonnegativity are unchanged. More generally, when the cost coefficients depend affinely on an uncertainty vector in a bounded polyhedron, dualizing the inner worst-case problem also gives a linear robust counterpart \citep{bertsimas2004price,bental2009robust}.

\subsection{Scenario weights and robust prices}\label{subsec:robust-dual-weights}

The dual of~\eqref{eq:robust-solver-primal} has one nonnegative multiplier $\omega_a$ for each scenario row, and these multipliers sum to one. They show which scenarios determine the robust value and the robust prices. Let $\Delta_M:=\{\bm{\omega}\in\R_+^M:\sum_{a=1}^M\omega_a=1\}$. For weights $\bm\omega\in\Delta_M$, define the mixed costs and their lower envelope across agents,
$$H_{ki}^{\bm{\omega}}:=\sum_{a=1}^M\omega_aH_{ki}^a,\qquad H_k^{\bm{\omega},\wedge}:=\min_{i\in[n]}H_{ki}^{\bm{\omega}},\qquad k=1,\dots,m.$$
The weights average the scenario costs, not the scenario probabilities. Even when all scenarios share the distortions $h_i$, $H_{ki}^{\bm\omega}$ in general differs from the cost $h_i\bigl(\sum_{a=1}^M\omega_a\bar P_k^a\bigr)$ under the mixed law $\sum_{a=1}^M\omega_a\P^a$, and it need not coincide with the costs of any candidate model. For concave distortions, $h_i\bigl(\sum_{a=1}^M\omega_a\bar P_k^a\bigr)\ge H_{ki}^{\bm\omega}$. A mixture of the candidate laws can then be assessed more severely than every candidate law, and the robust value over all such mixtures can exceed the robust value over $\Theta_M$.

\begin{corollary}
\label{cor:robust-lp-dual}
Program \eqref{eq:robust-solver-primal} has dual
\begin{equation}\label{eq:robust-solver-dual}
\begin{aligned}
(D_{\Pi}(\bm{b},\bm{r}_{\Lset};\Theta_M))\qquad
\max_{\bm{\lambda},\bm{\eta},\bm{\mu},\bm{\omega}}\quad
& \bm{s}^\top\bm{\lambda}
+\bm{r}_{\Lset}^\top\bm{\eta}
-\bm{b}^\top\bm{\mu}\\
\textnormal{s.t.}\quad
& \lambda_k
\le
H_{ki}^{\bm{\omega}}
-(E_{\Lset}^\top\bm{\eta})_{ki}
+
(A^\top\bm{\mu})_{ki},
&& k=1,\dots,m,\quad i\in[n],\\
& \bm{\mu}\ge\bm{0},\qquad
\bm{\omega}\in\Delta_M,\\
& \bm{\lambda}\in\R^m,\qquad
\bm{\eta}\in\R^{e_{\Lset}}.
\end{aligned}
\end{equation}
If $\mathcal C_\Pi(\Lset)$ is nonempty, strong duality holds, and every primal--dual optimum $(\bm x^\star,\tau^\star;\bm\lambda^\star,\bm\eta^\star,\bm\mu^\star,\bm\omega^\star)$ satisfies the following.
\begin{enumerate}[label=(\roman*)]
\item For every layer $k$ with $s_k>0$, $\lambda_k^\star=\min_{i\in[n]}\bigl(H_{ki}^{\bm\omega^\star}-(E_{\Lset}^\top\bm\eta^\star)_{ki}+(A^\top\bm\mu^\star)_{ki}\bigr)$.
\item If $x_{ki}^\star>0$, then $\lambda_k^\star=H_{ki}^{\bm\omega^\star}-(E_{\Lset}^\top\bm\eta^\star)_{ki}+(A^\top\bm\mu^\star)_{ki}$.
\item If $\omega_a^\star>0$, then scenario $a$ attains the robust value: $\sum_{k=1}^m\sum_{i=1}^nH_{ki}^ax_{ki}^\star=\tau^\star=V_\Pi(\bm b,\bm r_{\Lset};\Theta_M)$.
\end{enumerate}
\end{corollary}

Statements (i) and (ii) are the robust counterparts of the two identities in Theorem~\ref{thm:general-polyhedral-layer-pricing}: layers are priced as in the nominal problem, with the mixed costs $H_{ki}^{\bm\omega^\star}$ in place of $H_{ki}$. By (iii), a scenario whose total risk assessment of $\bm x^\star$ is strictly below the robust value has zero weight, so its costs do not enter the prices. The converse fails: a scenario may attain the robust value and still receive zero weight in a given dual optimum.

Without operational constraints, the result takes a particularly simple form, which extends Corollary~\ref{cor:layer_assignment} to several scenarios.

\begin{corollary}
\label{cor:robust-envelope-rule}
Suppose $\Lset_\Pi=\R^{mn}$. Then
\begin{equation}\label{eq:robust-unconstrained-envelope-value}
V_{\Pi}(\R^{mn};\Theta_M)
=
\max_{\bm{\omega}\in\Delta_M}\sum_{k=1}^m s_kH_k^{\bm{\omega},\wedge}.
\end{equation}
Let $\bm\omega^\star$ attain this maximum. A feasible allocation $\bm x$ is optimal for the robust program if and only if
\begin{enumerate}[label=(\alph*)]
\item for every layer $k$ with $s_k>0$, $x_{ki}>0$ only for agents $i$ with $H_{ki}^{\bm\omega^\star}=H_k^{\bm\omega^\star,\wedge}$, and
\item every scenario $a$ with $\omega_a^\star>0$ attains the largest total risk assessment of $\bm x$ across scenarios.
\end{enumerate}
\end{corollary}

Identity~\eqref{eq:robust-unconstrained-envelope-value} exchanges the minimum over allocations with the maximum over the scenario simplex. The robust planner thus behaves as if it faced a single scenario with the worst-case mixed costs $H_{ki}^{\bm\omega^\star}$, and each layer goes to an agent with the lowest mixed cost. If $\bm\omega^\star$ puts all its weight on one scenario, the robust allocation is a lower-envelope allocation for that scenario. If several scenarios have positive weight, condition (b) requires that they give the same total risk assessment, and the planner may need ties in the mixed costs to achieve this. A robust allocation may therefore have to split a layer between two agents, which is never needed in the unconstrained nominal problem. Section~\ref{subsec:cat-tower-regime} shows how the weights move between scenarios in the catastrophe tower.

With operational constraints, the same reasoning applies to shadow-adjusted mixed costs. For fixed $\bm\omega\in\Delta_M$, $\bm\mu\ge\bm0$ and $\bm\eta\in\R^{e_{\Lset}}$, define the robust adjusted envelope
\begin{equation}\label{eq:robust-shadow-adjusted-envelope}
H_k^{\bm{\omega},\bm{\eta},\bm{\mu};\Lset}
:=
\min_{i\in[n]}
\left(
H_{ki}^{\bm{\omega}}
-(E_{\Lset}^\top\bm{\eta})_{ki}
+(A^\top\bm{\mu})_{ki}
\right),
\qquad k=1,\dots,m.
\end{equation}
Choosing $\lambda_k$ at this envelope in the dual gives
\begin{equation}\label{eq:robust-dual-envelope-value}
V_{\Pi}(\bm{b},\bm{r}_{\Lset};\Theta_M)
=
\max_{\bm{\omega}\in\Delta_M,\ \bm{\mu}\ge\bm{0},\ \bm{\eta}\in\R^{e_{\Lset}}}
\left\{
\sum_{k=1}^m s_kH_k^{\bm{\omega},\bm{\eta},\bm{\mu};\Lset}
+\bm{r}_{\Lset}^\top\bm{\eta}
-\bm{b}^\top\bm{\mu}
\right\},
\end{equation}
which reduces to~\eqref{eq:robust-unconstrained-envelope-value} without operational rows. The dual feasible set in~\eqref{eq:robust-solver-dual} does not depend on $(\bm b,\bm r_{\Lset})$. Hence Corollary~\ref{cor:operational-shadow-values} and Proposition~\ref{prop:one-sided-sensitivity} hold for the robust value with the same proofs: $V_{\Pi}(\bm{b},\bm{r}_{\Lset};\Theta_M)$ is convex and polyhedral, nonincreasing in $\bm b$, with subgradient $(-\bm\mu^\star,\bm\eta^\star)$, and its one-sided derivatives are computed from the robust optimal multipliers. Section~\ref{subsec:capacity-sensitivity} applies this to capacity limits.

The layer prices $\bm{\lambda}^{\star}$ price any loss payment $Y=f(S)$ that is nondecreasing in the aggregate loss, such as an agent's allocated loss, through the functional $\mathsf{P}_{\bm{\lambda}^{\star}}$ of Section~\ref{ss:constrained-price-system}. As in the nominal problem, $\mathsf{P}_{\bm{\lambda}^{\star}}(Y)=\E^{\Q_{\lambda^{\star}}}[Y]$ when $\bm{\lambda}^{\star}$ is normalized and nonincreasing. This holds without operational constraints, because $k\mapsto H_{ki}^{\bm\omega}$ is nonincreasing with $H_{1i}^{\bm\omega}=1$; with capacity limits only, the argument of Corollary~\ref{cor:scarcity-adjusted-pricing} applies to the mixed costs. Finally, the robust dual may have several optimal solutions, for instance several worst-case weight vectors. They all support the same optimal allocations but may give different layer prices, and the other multipliers matter only through $\bm\lambda$. The supporting price of a payment $Y$ is then any value between the infimum and the supremum of $\mathsf{P}_{\bm\lambda}(Y)$ over the optimal dual solutions, and it is unique only when the two coincide.

\section{Application to capacity caps}\label{sec:caps}

We now specialize Section~\ref{sec:main}'s constrained program to capacity caps, which place a limit on the total loss an agent may absorb. Such limits are important because finite capital and risk-bearing ability restrict the exposure an insurer can bear. They may reflect capital requirements, contractual commitments, or internal limits. Since cumulative allocations are nondecreasing in the aggregate loss, it suffices to impose the capacity constraint at the final grid state:
$$X_{mi}=\sum_{k=1}^m x_{ki}\le C_i.$$
The terminal inequality is equivalent to $X_i\le C_i$ almost surely.

With these caps, allocation becomes a classical transportation problem. Layer $k$ contains $s_k$ units of loss to be allocated, and agent $i$ can absorb at most $C_i$ units in total. The program is therefore feasible if and only if the agents' combined capacity is at least the maximum aggregate loss, that is, $\sum_{i=1}^n C_i\ge S_m$. The primal distributes all layer mass within those budgets. Its capacity multipliers are interpreted as scarcity prices, as developed in Section~\ref{subsec:capacity-dual-scarcity}. \citet{PeyreCuturi2020} give a modern account of discrete transportation and its duality.

\subsection{Capacity program and scarcity prices}
\label{subsec:capacity-dual-scarcity}

Treat $C_i$ as a budget on the total loss increments assigned to agent $i\in[n]$. In layer variables, the resulting transportation program is
\begin{equation}\label{eq:capped_transport_nonneg}
\begin{aligned}
\min_{\bm{x}\ge\bm{0}}\quad &\sum_{k=1}^m\sum_{i=1}^n H_{ki}x_{ki}\\
\textnormal{s.t.}\quad
&\sum_{i=1}^n x_{ki}=s_k, \qquad k=1,\dots,m,\\
&\sum_{k=1}^m x_{ki}\le C_i, \qquad i\in[n].
\end{aligned}
\end{equation}

Program \eqref{eq:capped_transport_nonneg} can be solved by generic linear-programming methods or by transportation routines that exploit its bipartite flow structure, including network simplex \citep{ahuja1993network} and successive shortest augmenting paths \citep{ford1962flows}.

A capacity row for agent $i$ has coefficient one on $x_{ki}$ for every $k$, so its multiplier $\mu_i\ge0$ enters the dual through $(A^\top\bm{\mu})_{ki}=\mu_i$, and the agent's adjusted layer cost is $H_{ki}+\mu_i$. In minimum cost flow terminology, $\lambda_k$ and $\mu_i$ are the node potentials for layer and agent nodes. The scarcity interpretation dates to \citet{koopmans1949optimum}, and complementary slackness gives the reduced cost optimality conditions for the network simplex method \citep[see][Chapter~11]{ahuja1993network}. Nonnegative capacities satisfying $\sum_{i=1}^n C_i\ge S_m$ make \eqref{eq:capped_transport_nonneg} feasible, and its value is finite because $0\le H_{ki}\le1$. The following proposition gives the dual program.

\begin{proposition}\label{prop:capped-transport-dual}
Suppose $C_i\ge0$ for $i\in[n]$ and $\sum_{i=1}^n C_i\ge S_m$. Then \eqref{eq:capped_transport_nonneg} has dual
\begin{equation}\label{eq:capped_layer_transport_dual}
\begin{aligned}
\max_{\bm{\lambda}\in\R^m,\ \bm{\mu}\in\R_+^n}\quad&
\sum_{k=1}^m \lambda_k s_k - \sum_{i=1}^n \mu_i C_i\\
\textnormal{s.t.}\quad&
\lambda_k \le H_{ki}+\mu_i,\quad k=1,\dots,m,\ i\in[n].
\end{aligned}
\end{equation}
For any optimal pair $(\bm{x}^C,(\bm{\lambda}^C,\bm{\mu}^C))$, each layer $k=1,\dots,m$ with $s_k>0$ satisfies
$$\lambda_k^C=\min_{i\in[n]}(H_{ki}+\mu_i^C).$$
For every $k=1,\dots,m$ and $i\in[n]$, a positive assignment $x_{ki}^C>0$ implies $\lambda_k^C=H_{ki}+\mu_i^C$. Finally, $\mu_i^C>0$ implies $\sum_{k=1}^m x_{ki}^C=C_i$ for each $i\in[n]$, so agent $i$'s cap binds.
\end{proposition}

If agent $i$'s capacity has a positive multiplier $\mu_i^C$, assigning a layer to that agent uses capacity that could otherwise cover another layer. We account for this cost by adding $\mu_i^C$ to every distortion evaluation $H_{ki}$. The resulting curve is shifted upward by the same amount on every layer. We then take the lower envelope of these adjusted curves, $\lambda_k^C=\min_{i\in[n]}(H_{ki}+\mu_i^C)$, and assign each layer only to agents attaining that minimum.

If $\sum_{i=1}^n C_i=S_m$, the dual solution is not unique. Adding a constant $c$ to all multipliers $\lambda_k^C$, $k=1,\dots,m$, and $\mu_i^C$, $i=1,\dots,n$, while keeping $\mu_i^C+c\ge0$ for every $i\in[n]$, leaves every dual constraint unchanged. It also leaves the objective unchanged because $\sum_{k=1}^m s_k=\sum_{i=1}^n C_i$. We remove this common-shift freedom by subtracting $\min_{i\in[n]}\mu_i^C$ from both vectors, selecting $\min_{i\in[n]}\mu_i^C=0$. This normalization gives $\lambda_1^C=1$, which ensures that the pricing masses sum to one. If $\sum_{i=1}^n C_i>S_m$, at least one agent is assigned less than its capacity at an optimum. Complementary slackness sets that agent's capacity multiplier to zero, so the normalization holds automatically.

\begin{corollary}\label{cor:scarcity-adjusted-pricing}
For \eqref{eq:capped_transport_nonneg}, choose an optimal dual representative $(\bm{\lambda}^C,\bm{\mu}^C)$ with $\min_{i\in[n]}\mu_i^C=0$, and, if $s_1=0$, with $\lambda_1^C$ selected as in Remark~\ref{rem:zero-size-first-layer}. Its constrained prices obey
$$\lambda_k^C=\min_{i\in[n]}(H_{ki}+\mu_i^C), \qquad k=1,\dots,m.$$
The measure $\Q^C$ defined by
$$\Q^C(A_k):=q_k^C:=q_k(\bm{\lambda}^C), \qquad k=1,\dots,m,$$
with the convention in \eqref{eq:pricing-mass-convention}, is a probability measure.
\end{corollary}

If every cap is slack at a primal optimum, complementary slackness sets $\bm{\mu}^C=\bm{0}$ in every paired dual optimum. Prices then collapse to the unconstrained envelope $\lambda_k^\star=H_k^\wedge$ from Corollary~\ref{cor:dual_layer_prices}.

\subsection{Capacity sensitivity}\label{subsec:capacity-sensitivity}

We now specialize Proposition~\ref{prop:one-sided-sensitivity} to capacity limits, which we parameterize by the capacity vector. Fix a set of agents $J\subseteq[n]$ who have capacity constraints, and for $\bm{C}=(C_i)_{i\in J}\in\R_+^J$ define
\begin{equation}\label{eq:parametric-capacity-value}
\begin{aligned}
v_J(\bm{C}):=
\min_{\bm{x}\ge\bm{0}}\quad
&\sum_{k=1}^m\sum_{i=1}^n H_{ki}x_{ki}\\
\textnormal{s.t.}\quad
&\sum_{i=1}^n x_{ki}=s_k, \qquad k=1,\dots,m,\\
&\sum_{k=1}^m x_{ki}\le C_i, \qquad i\in J.
\end{aligned}
\end{equation}
Agents outside $J$ remain uncapped. The domain of $v_J$ is the set of capacity vectors for which the program has a feasible allocation and hence a finite optimal value:
$$\operatorname{dom}v_J:=\{\bm{C}\in\R_+^J:v_J(\bm{C})<\infty\}.$$
Collect the variables satisfying the dual constraints in
$$\mathcal Z_J^{\rm cap} := \left\{ (\bm{\lambda},\bm{\mu})\in\R^m\times\R_+^J:
\begin{aligned}
&\lambda_k\le H_{ki}+\mu_i && k=1,\dots,m,\ i\in J,\\
&\lambda_k\le H_{ki} && k=1,\dots,m,\ i\notin J
\end{aligned}
\right\}.$$
At $\bm{C}\in\operatorname{dom}v_J$, let
$$\mathcal Z_J^{\rm cap,\star}(\bm{C}) := \operatorname*{arg\,max}_{\displaystyle(\bm{\lambda},\bm{\mu})\in\mathcal Z_J^{\rm cap}} \left( \sum_{k=1}^m\lambda_ks_k-\sum_{i\in J}\mu_iC_i \right)$$
be the dual optimizer set. Its projection onto capacity multipliers is
$$\mathcal M_J(\bm{C}):=\left\{\bm{\mu}:(\bm{\lambda},\bm{\mu})\in\mathcal Z_J^{\rm cap,\star}(\bm{C})\ \text{for some}\ \bm{\lambda}\right\},$$
so its elements are the capacity-multiplier vectors from all optimal dual solutions. The set can contain several vectors when those solutions have different capacity multipliers. Since $v_J$ is the value function of a program of the form $P_\Pi(\bm b)$, with the capacity rows of the agents in $J$ as operational inequalities and $\bm b=\bm C$, Corollary~\ref{cor:operational-shadow-values} and linear-programming sensitivity analysis \citep[Section~5.2]{bertsimas1997introduction} show that $v_J$ is coordinatewise nonincreasing, convex, and polyhedral on $\operatorname{dom}v_J$, with $-\bm\mu\in\partial v_J(\bm C)$ for every $\bm\mu\in\mathcal M_J(\bm C)$. Proposition~\ref{prop:one-sided-sensitivity}, with $\mathcal M_J(\bm C)$ in the role of $\mathcal M(\bm b)$, gives its one-sided derivatives.

The relation between a capacity limit and its marginal value is easier to see when we vary one agent's capacity and keep the others fixed.

\begin{corollary}\label{cor:single-capacity-value}
Fix an agent $i\in J$ and keep the capacities of all other agents fixed. Let $\bm C(\kappa)$ set agent $i$'s capacity to $\kappa$. On an interval containing $0$ where $\bm{C}(\kappa)\in\operatorname{dom}v_J$, define
$$G(\kappa):=v_J(\bm{C}(0))-v_J(\bm{C}(\kappa)),\qquad \mathcal M_i(\kappa):=\mathcal M_J(\bm{C}(\kappa)).$$
Then
$$G'(\kappa+)=\min_{\bm{\mu}\in\mathcal M_i(\kappa)}\mu_i.$$
If $\kappa$ is an interior point of that interval (the left derivative exists by convexity of $v_J$), then
$$G'(\kappa-)=\max_{\bm{\mu}\in\mathcal M_i(\kappa)}\mu_i.$$
Thus $G$ is differentiable at $\kappa$ exactly when $\mu_i$ is constant on $\mathcal M_i(\kappa)$. In that event, every optimal dual pair gives $G'(\kappa)=\mu_i$ and $\frac{d}{d\kappa}v_J(\bm{C}(\kappa))=-\mu_i$.
\end{corollary}

To measure the benefit of expansion at a breakpoint, the relevant multiplier is therefore the smallest optimal one rather than an arbitrary multiplier returned by the solver.

The same sensitivity results hold for the robust program (Section~\ref{subsec:robust-dual-weights}). Let $v_J^{\rm rob}(\bm C)$ denote the robust value when the operational rows are the capacity rows of the agents in $J$, and let $\mathcal M_J^{\rm rob}(\bm C)$ collect the capacity multipliers of its optimal dual solutions. Then $v_J^{\rm rob}$ is coordinatewise nonincreasing, convex, and polyhedral, and Corollary~\ref{cor:single-capacity-value} holds with $v_J$ and $\mathcal M_J$ replaced by $v_J^{\rm rob}$ and $\mathcal M_J^{\rm rob}$. Setting $\mu_i:=0$ for $i\notin J$, the capacity rows give $(A^\top\bm{\mu})_{ki}=\mu_i$ in \eqref{eq:robust-shadow-adjusted-envelope}. For $s_k>0$, the robust layer price at a primal--dual optimum with dual variables $(\bm\lambda,\bm\mu,\bm\omega)$ is therefore $\lambda_k=\min_{i\in[n]}(H_{ki}^{\bm{\omega}}+\mu_i)$: scenarios are mixed first, and the capacity adjustments are applied next.

\subsection{Greedy rule with one constrained agent}\label{subsec:greedy-one-cap}

\begin{figure}[ht]
\centering
\includegraphics[width=\textwidth]{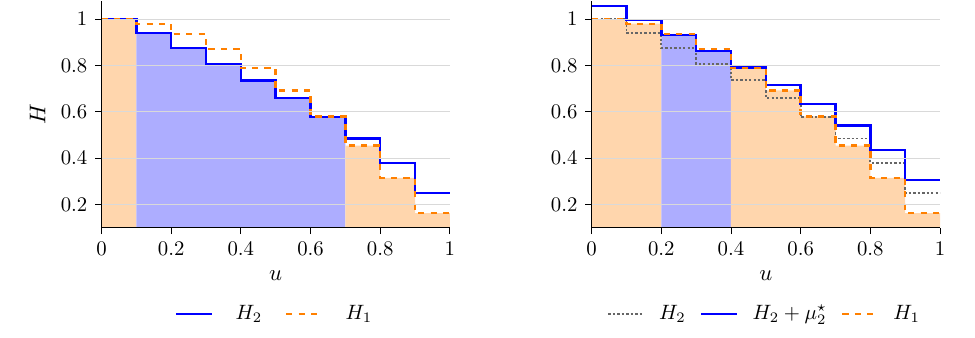}
\caption{Greedy rule with one capacity-constrained agent, illustrated with two agents. The horizontal axis is the probability level $u$, and layer $k$ occupies $[P_{k-1},P_k)$. Left: unconstrained lower envelope of the cumulative costs. Right: envelope after the capped agent's curve is shifted up by its scarcity price $\mu_n^C$. Shading indicates which agent is assigned each loss layer.}
\label{fig:one-cap-greedy}
\end{figure}

When only one agent is capped, \eqref{eq:capped_transport_nonneg} is solved by a simple greedy rule. Without the cap, every layer goes to its cheapest agent. If the capped agent has the lowest distortion evaluation on more loss than it can absorb, one can lift its cost curve by a constant until it can be assigned exactly its capacity from layers on which it has the lowest adjusted evaluation, splitting tied layers when necessary. Each remaining layer can be assigned to an uncapped agent with the lowest evaluation on that layer. The amount by which the distortion evaluations are lifted is the scarcity price $\mu_n^C$ from Proposition~\ref{prop:capped-transport-dual}. Figure~\ref{fig:one-cap-greedy} shows the rule for two agents: the left panel is the unconstrained lower envelope, and the right panel is the envelope after the capped agent's curve is shifted up by $\mu_n^C$. The shaded areas identify which agent is assigned each loss layer.

Let $\overline H_k:=\min_{i\in[n-1]}H_{ki}$ be the cheapest uncapped cost on layer $k$, and let $\Delta_k:=\overline H_k-H_{kn}$ be agent~$n$'s comparative advantage on that layer. After lifting agent~$n$'s distortion curve by $\mu$, agent~$n$ is strictly cheapest on layers with $\Delta_k>\mu$ and ties the cheapest uncapped agent when $\Delta_k=\mu$. Lifting until the cap holds is therefore the same as filling layers in decreasing order of $\Delta_k$ until capacity runs out. We collect this procedure in Algorithm~\ref{alg:one-cap-greedy}. The algorithm assigns no layer with $\Delta_k\le0$ to agent~$n$. On a layer with $\Delta_k=0$, agent~$n$ ties the cheapest uncapped agent, so moving that layer to agent~$n$, as far as its remaining capacity permits, gives an alternative optimal allocation with the same value.

\begin{algorithm}[H]
\caption{Greedy allocation when a single agent is capacity constrained}
\label{alg:one-cap-greedy}
\begin{algorithmic}[1]
\Require Layer sizes $s_1,\dots,s_m$, costs $H_{ki}$, and capacity $C\ge0$ of agent~$n$
\State $x_{ki}\gets0$ for all $k=1,\dots,m$ and $i\in[n]$; $c\gets C$
\State $\Delta_k\gets\min_{i\in[n-1]}H_{ki}-H_{kn}$ for $k=1,\dots,m$
\For{$k$ in decreasing order of $\Delta_k$, while $\Delta_k>0$}
  \State $x_{kn}\gets\min\{s_k,c\}$; $c\gets c-x_{kn}$
\EndFor
\State Give the remaining mass $s_k-x_{kn}$ of every layer to an agent in $\operatorname*{arg\,min}_{i\in[n-1]}H_{ki}$
\State \Return $\bm x$
\end{algorithmic}
\end{algorithm}

\begin{proposition}\label{prop:one-cap-greedy}
Let $n\ge2$, suppose agent $n$ has capacity $C_n=C\ge0$, and assume $C_i\ge S_m$ for all $i\in[n-1]$. Algorithm~\ref{alg:one-cap-greedy} then returns an optimal solution of \eqref{eq:capped_transport_nonneg}, whose value is the constrained planner value.
\end{proposition}

Since the uncapped agents can absorb any remaining loss, the total cost of the allocation is $\sum_{k=1}^m\overline H_ks_k-\sum_{k=1}^m\Delta_kx_{kn}$. The program therefore reduces to the continuous knapsack problem of choosing $0\le x_{kn}\le s_k$ with $\sum_{k=1}^mx_{kn}\le C$ to maximize the saving $\sum_{k=1}^m\Delta_kx_{kn}$. Algorithm~\ref{alg:one-cap-greedy} is the classical greedy solution of this problem \citep{dantzig1957discrete} and needs $O(mn+m\log m)$ operations. In continuous models, \citet{lo2017neyman} obtains threshold rules of the same type for constrained reinsurance problems under distortion risk measures.

The dual solution constructed in the proof of Proposition~\ref{prop:one-cap-greedy} (Appendix~\ref{sec:proofs}) also gives the marginal value of additional capacity for agent~$n$. If capacity runs out inside a layer $k$, then $\mu_n=\Delta_{k}$ is the unique optimal multiplier, and Corollary~\ref{cor:single-capacity-value} gives $G'(\kappa)=\Delta_{k}$: one more unit of capacity is worth agent~$n$'s comparative advantage on the layer it is currently splitting. If capacity runs out exactly at the end of layer $k$, the same complementary-slackness argument shows that the optimal multipliers form the interval from the largest positive advantage among the remaining layers, or zero if none remains, up to $\Delta_{k}$, and these endpoints are $G'(\kappa+)$ and $G'(\kappa-)$.

\section{Catastrophe tower case study}\label{sec:numerical-examples}

We now apply our method to a stylized catastrophe reinsurance tower. Five participants share an aggregate catastrophe loss: a primary risk-sharing pool, two reinsurers, a catastrophe-bond tranche and a public backstop. All participants except the pool have limited capacity, the catastrophe bond and the backstop can only cover given loss bands, the two reinsurers share a common cession limit, and the pool must retain part of each layer below the catastrophe-bond attachment. These constraints link different layers and different participants, so the layerwise rule of Theorem~\ref{thm:llw_th3} no longer applies, and we solve the linear program~\eqref{eq:planner-grid} instead. We first describe the example (Section~\ref{subsec:cat-tower-setup}) and its optimal allocation (Section~\ref{subsec:cat-tower-tower}). We then compute the value of additional public capacity (Section~\ref{subsec:cat-tower-demand}) and study how uncertainty about the catastrophe model changes this value (Section~\ref{subsec:cat-tower-regime}).

\subsection{Setup}\label{subsec:cat-tower-setup}

We consider the annual aggregate loss, measured in billions. We divide the range $[0,250]$ of the aggregate loss into $m=2{,}500$ layers of width $0.1$, and losses above $250$ are set to $250$. A participant's share of a layer is the fraction of that layer's loss that it pays. For instance, a participant that holds the whole band $(90,130]$ pays nothing when the aggregate loss $S$ is below $90$, pays $S-90$ when $S$ lies between $90$ and $130$, and pays $40$ when $S$ exceeds $130$.

We consider three probability scenarios. Scenarios $\P^{\mathrm A}$ and $\P^{\mathrm B}$ represent two vendor catastrophe models, and $\P^{\mathrm A}$ is the nominal model. The climate-conditioned scenario $\P^{\mathrm{CL}}$ has a higher event frequency than $\P^{\mathrm A}$ and a heavier severity tail. In each scenario, a Poisson number of events occurs and each event causes an independent lognormal loss; we compute the distribution of the aggregate loss by fast Fourier transform. Table~\ref{tab:cat-tower-scenarios} reports the parameters, the mean aggregate loss, and the probabilities of exceeding $90$, $130$ and $175$, which are the endpoints of the bands used below.

\begin{table}[ht]
\centering
\small
\setlength{\tabcolsep}{4pt}
\caption{Compound-Poisson scenarios: inputs, mean aggregate loss (before losses above $250$ are set to $250$), and probabilities of exceeding the band endpoints.}
\label{tab:cat-tower-scenarios}
\begin{tabular}{lcccccc}
\toprule
Scenario & Frequency & $\log$ severity & Mean loss & $\P(S>90)$ & $\P(S>130)$ & $\P(S>175)$\\
\midrule
$\P^{\mathrm A}$ nominal vendor & $3.00$ & $(\log 1.50,\ 1.45)$ & $12.9$ & $0.0102$ & $0.0042$ & $0.0019$\\
$\P^{\mathrm B}$ second vendor & $4.75$ & $(\log 2.20,\ 1.15)$ & $20.2$ & $0.0080$ & $0.0019$ & $0.0006$\\
$\P^{\mathrm{CL}}$ climate-conditioned & $3.60$ & $(\log 1.55,\ 1.52)$ & $17.7$ & $0.0207$ & $0.0090$ & $0.0044$\\
\bottomrule
\end{tabular}
\end{table}

Scenario $\P^{\mathrm B}$ has the largest mean loss but the thinnest tail. The backstop attachment $130$ is about a 1-in-240-year loss under $\P^{\mathrm A}$, a 1-in-520-year loss under $\P^{\mathrm B}$, and a 1-in-110-year loss under $\P^{\mathrm{CL}}$. The distribution function of $\P^{\mathrm{CL}}$ lies below that of $\P^{\mathrm A}$ at every loss level, so $\P^{\mathrm{CL}}$ first-order stochastically dominates $\P^{\mathrm A}$. The distribution functions of $\P^{\mathrm A}$ and $\P^{\mathrm B}$ cross at a loss of about $75$, and neither vendor model dominates the other.

Table~\ref{tab:cat-tower-agents} describes the participants. We chose the parameters for illustration; they are not calibrated to an actual insurance program. Under the nominal model, the catastrophe bond attaches at about the 1-in-100-year loss and both bands end at about the 1-in-500-year loss.

\begin{table}[!ht]
\centering
\caption{Participants and constraints of the catastrophe tower.}
\label{tab:cat-tower-agents}
\begin{tabular}{llccc}
\toprule
Agent & Role & Distortion $h_i(p)$ & Capacity & Eligible band\\
\midrule
$P$ & primary risk-sharing pool & $0.92p+0.08\min\{p/0.01,1\}$ & uncapped & $[0,250]$\\
$\mathrm{R1}$ & first reinsurer & $p^{0.90}$ & $30$ & $[0,250]$\\
$\mathrm{R2}$ & second reinsurer & $p^{0.86}$ & $30$ & $[0,250]$\\
$\mathrm{ILS}$ & catastrophe-bond tranche & $\min\{1.65p,1\}$ & $50$ & $(90,175]$\\
$\mathrm{GOV}$ & public backstop & $\min\{1.15p,1\}$ & $C_{\mathrm{GOV}}$ & $(130,175]$\\
\midrule
\multicolumn{5}{l}{Retention corridor: $x_{kP}\ge0.10\,s_k$ for every layer $k$ with $S_k\le90$.}\\
\multicolumn{5}{l}{Cession bound: $\sum_{k=1}^m\left(x_{k,\mathrm{R1}}+x_{k,\mathrm{R2}}\right)\le52$.}\\
\bottomrule
\end{tabular}
\end{table}

The pool evaluates its loss with a weighted sum of the expected loss (weight $92\%$) and the Expected Shortfall at level $99\%$ (weight $8\%$). It is close to risk neutral for frequent losses, but the Expected Shortfall part makes rare losses costly to retain: on a layer whose exceedance probability $p$ is below $0.01$, the pool's cost per unit is $8.92p$, almost nine times the expected loss. Both reinsurers use proportional-hazard distortions and have capacity $30$. The second reinsurer has the smaller exponent, so it is more expensive than the first on every layer and is only useful once the first is full. The catastrophe bond and the backstop charge the expected loss plus a loading of $65\%$ and $15\%$, respectively, so their costs per unit in their bands are $1.65p$ and $1.15p$.

Two constraints, listed at the bottom of Table~\ref{tab:cat-tower-agents}, link several layers or participants. The retention corridor requires the pool to keep at least $10\%$ of each layer below the catastrophe-bond attachment. The cession bound limits the total amount ceded to the two reinsurers to $52$, below their combined capacity of $60$; the catastrophe bond and the backstop do not count toward this bound. Because the cession bound involves two participants, the program is not a transportation problem, and we use the general prices of Theorem~\ref{thm:general-polyhedral-layer-pricing} rather than Proposition~\ref{prop:capped-transport-dual}. We write $\mu_{\mathrm{R1}}$, $\mu_{\mathrm{R2}}$, $\mu_{\mathrm{ILS}}$ and $\mu_{\mathrm{GOV}}$ for the multipliers of the capacity limits and $\mu_{\mathrm{off}}$ for the multiplier of the cession bound. We solve the linear programs with the HiGHS solver \citep{huangfu2018parallelizing} in R. The nominal program has $8{,}800$ variables, and the whole study, about $1{,}500$ linear programs, runs in a few seconds.

\begin{figure}[ht]
\centering
\includegraphics[width=\textwidth]{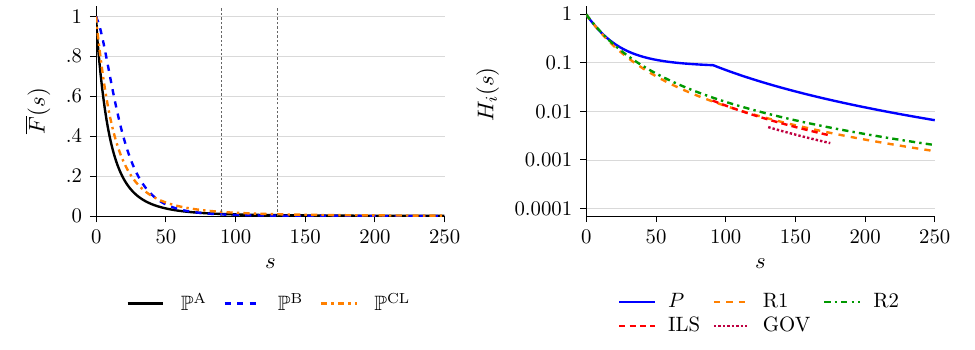}
\caption{Left: exceedance probabilities $\bar F(s)=\P(S>s)$ under the three scenarios; the vertical lines mark $90$ and $130$. Right: nominal costs $H_i(s)=h_i\bigl(\P^{\mathrm A}(S>s)\bigr)$ of the five participants (log scale).}
\label{fig:cat-tower-instance}
\end{figure}

Figure~\ref{fig:cat-tower-instance} shows the exceedance probabilities under the three scenarios and the nominal costs $H_{ki}=h_i(\bar P_k)$ of the five participants. In the backstop's band $(130,175]$, the backstop is the cheapest participant, followed by the catastrophe bond and the first reinsurer, and the pool is by far the most expensive. The kink in the pool's cost near $90$ is where the nominal exceedance probability falls below $1\%$. These cost differences make it valuable to move tail losses out of the pool, and the constraints determine how much of this value can be obtained.

\subsection{The optimal tower}\label{subsec:cat-tower-tower}

As a benchmark, we first remove the capacity limits, the corridor and the cession bound, and keep only the eligible bands. Each layer then goes to its cheapest eligible participant (Corollary~\ref{cor:layer_assignment}). Under the nominal model, the pool keeps the layers below $4.6$, the first reinsurer takes $(4.6,107.4]$ and the layers above $175$, the catastrophe bond takes $(107.4,130]$, and the backstop takes $(130,175]$. The minimum total risk assessment is $14.84$. Since removing constraints can only lower the minimum, this is a lower bound for all the constrained problems below.

\begin{figure}[ht]
\centering
\includegraphics[width=\textwidth]{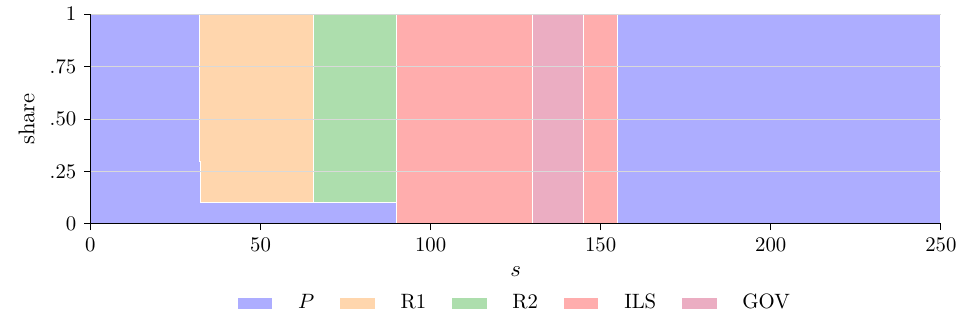}
\caption{Optimal nominal allocation at $C_{\mathrm{GOV}}=15$: share of each layer held by each participant.}
\label{fig:cat-tower-allocation}
\end{figure}

Figure~\ref{fig:cat-tower-allocation} shows the optimal nominal allocation with all the constraints and public capacity $C_{\mathrm{GOV}}=15$. From the bottom to the top of the tower:
\begin{itemize}[noitemsep]
\item the pool keeps all losses up to about $32$;
\item between $32$ and $90$, the pool keeps the required $10\%$ of each layer and the reinsurers take the other $90\%$, the first reinsurer up to $65.5$ and the second reinsurer above;
\item the catastrophe bond covers $(90,130]$;
\item the backstop covers $(130,145]$, which uses all of its capacity;
\item the catastrophe bond covers $(145,155]$ with its remaining capacity;
\item the pool keeps all losses above $155$.
\end{itemize}
The minimum total risk assessment is now $17.05$.

The constraints explain the differences from the benchmark. The reinsurers take $52$ in total, the maximum allowed by the cession bound, with $30$ for the first reinsurer and $22$ for the second, so the second reinsurer is not full. They place these $52$ units between $32$ and $90$, where replacing the pool saves the most, and no longer cover the layers above $175$. The catastrophe bond is split in two parts because the backstop is cheaper on $(130,145]$; the capacity that the catastrophe bond does not use there goes to $(145,155]$, where it replaces the pool. Above $155$, the catastrophe bond and the backstop have no capacity left, and the pool keeps the far tail although it is the most expensive participant there.

The multipliers show which constraints matter at the margin. At $C_{\mathrm{GOV}}=15$, the cession bound has the largest multiplier among the capacity and cession constraints, $\mu_{\mathrm{off}}=0.044$. Letting the reinsurers take one more unit would lower the total risk assessment by about $0.044$. The first reinsurer's capacity is worth much less, $\mu_{\mathrm{R1}}=0.005$, because while the cession bound binds, extra capacity at the first reinsurer can only replace the second reinsurer, which is slightly more expensive. The second reinsurer is not full, so $\mu_{\mathrm{R2}}=0$. The catastrophe bond and the backstop have multipliers $\mu_{\mathrm{ILS}}=0.019$ and $\mu_{\mathrm{GOV}}=0.021$; Section~\ref{subsec:cat-tower-demand} explains these two values.

\begin{figure}[ht]
\centering
\includegraphics[width=\textwidth]{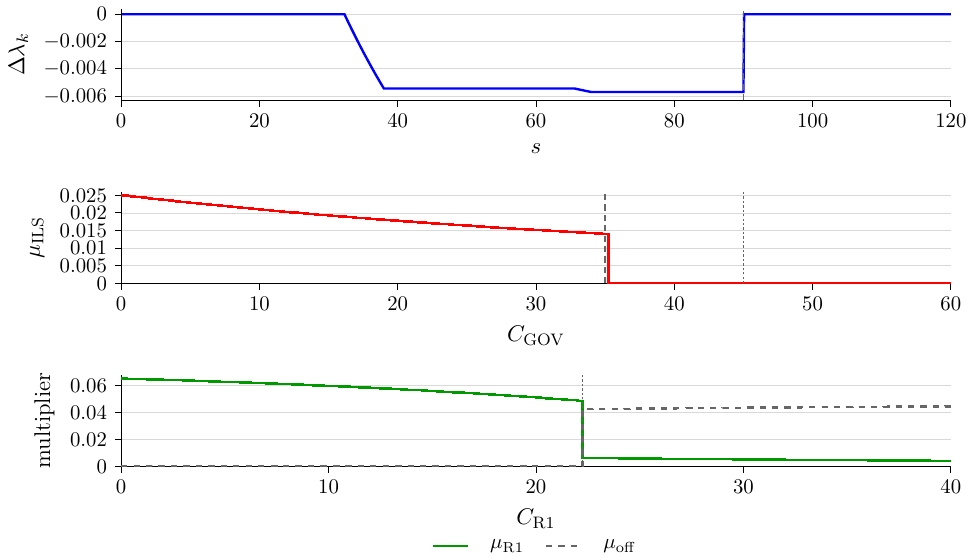}
\caption{Nominal model. Top: change in the layer prices $\lambda_k$ when the retention corridor is imposed. Middle: multiplier $\mu_{\mathrm{ILS}}$ of the catastrophe-bond capacity as public capacity varies. Bottom: multipliers $\mu_{\mathrm{R1}}$ and $\mu_{\mathrm{off}}$ as the first reinsurer's capacity varies ($C_{\mathrm{GOV}}=5$).}
\label{fig:cat-tower-incidence}
\end{figure}

The bottom panel of Figure~\ref{fig:cat-tower-incidence} shows how $\mu_{\mathrm{R1}}$ and $\mu_{\mathrm{off}}$ change when the first reinsurer's capacity $C_{\mathrm{R1}}$ varies from $0$ to $40$. (The panel uses $C_{\mathrm{GOV}}=5$, but public capacity does not affect the layers below $90$.) For $C_{\mathrm{R1}}<22$, both reinsurers are full but hold less than $52$ together, so the cession bound is slack. An extra unit of capacity at the first reinsurer then replaces the pool, and $\mu_{\mathrm{R1}}$ is large, falling from $0.065$ at $C_{\mathrm{R1}}=0$ to $0.049$ just below $22$. At $C_{\mathrm{R1}}=22=52-30$, the cession bound starts to bind. From then on, extra capacity at the first reinsurer can only replace the second reinsurer, so $\mu_{\mathrm{R1}}$ drops to $0.006$, while $\mu_{\mathrm{off}}$ jumps from $0$ to $0.043$. To move more risk out of the pool, the cession bound itself must be relaxed.

The corridor has a cost, and it also changes the supporting layer prices. We measure both by solving the nominal program again without the corridor. Removing the corridor lowers the minimum total risk assessment by $0.080$, whatever the public capacity, because the corridor only concerns layers below $90$ while public capacity only changes the allocation above $90$. Without the corridor, the reinsurers would take whole layers on $(38,90]$. With the corridor they can only take $90\%$ of each layer, so their $52$ units extend down to $32$. The last unit of reinsurance then replaces the pool on a lower layer, where it saves less, and the cession multiplier falls from $0.049$ to $0.044$. This multiplier enters the reinsurers' adjusted costs in Theorem~\ref{thm:general-polyhedral-layer-pricing}, so the supporting prices of the layers between $32$ and $90$ fall, by up to $0.0057$ (top panel of Figure~\ref{fig:cat-tower-incidence}); prices below $32$ and above $90$ do not change. On these layers the pool holds $10\%$ at a price below its own cost $H_{kP}$. Theorem~\ref{thm:general-polyhedral-layer-pricing} allows this, since the multiplier of the lower bound $x_{kP}\ge0.10\,s_k$ enters the pool's adjusted cost with a negative sign. Hence the corridor raises the cost of the program but lowers the supporting prices of the layers on which it binds.

\subsection{Demand curve for public capacity}\label{subsec:cat-tower-demand}

\begin{figure}[ht]
\centering
\includegraphics[width=\textwidth]{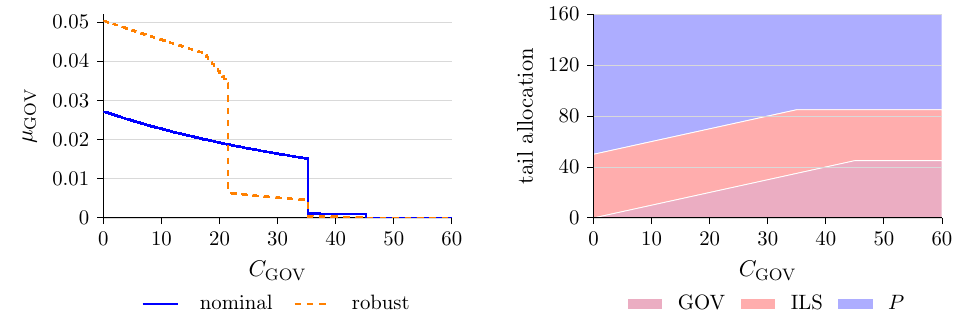}
\caption{Left: demand curves $\mu_{\mathrm{GOV}}$ for public capacity under the nominal and the robust programs. Right: nominal allocation of the losses above $90$ between the backstop, the catastrophe bond and the pool.}
\label{fig:cat-tower-demand}
\end{figure}

We now vary the public capacity $C_{\mathrm{GOV}}$ from $0$ to $60$ in steps of $0.25$, keep all other constraints fixed, and solve the nominal and the robust programs at each value. The multiplier $\mu_{\mathrm{GOV}}(C_{\mathrm{GOV}})$ of the backstop's capacity is the reduction in the minimum total risk assessment from one more unit of public capacity, after the whole tower has been re-optimized. We call $C_{\mathrm{GOV}}\mapsto\mu_{\mathrm{GOV}}(C_{\mathrm{GOV}})$ the demand curve for public capacity, because it gives the benefit to compare with the cost of an additional unit of capacity. The minimum total risk assessment is convex and piecewise linear in $C_{\mathrm{GOV}}$. At a kink, the smallest optimal multiplier is the saving from an expansion and the largest is the cost of a contraction (Proposition~\ref{prop:one-sided-sensitivity} and Corollary~\ref{cor:single-capacity-value}). We report the multiplier returned by the solver. In this example, the two one-sided values differ only slightly, except where the curves jump. Figure~\ref{fig:cat-tower-demand} shows the demand curves (left) and how the nominal allocation of the losses above $90$ changes with $C_{\mathrm{GOV}}$ (right). We discuss the nominal curve here and the robust curve in Section~\ref{subsec:cat-tower-regime}.

Consider first $C_{\mathrm{GOV}}=0$. The catastrophe bond then covers $(90,140]$, and the pool keeps all losses above $140$. A first unit of public capacity does not replace the pool directly. It replaces the catastrophe bond at $130$, the bottom of the backstop's band, and the catastrophe bond uses the freed unit at $140$, where it replaces the pool. Write $p(x):=\P^{\mathrm A}(S>x)$. Above $130$, $p(x)<0.01$, and the costs per unit are $8.92\,p(x)$ for the pool, $1.65\,p(x)$ for the catastrophe bond and $1.15\,p(x)$ for the backstop. The two substitutions save
$$\mu_{\mathrm{GOV}}(0)=(1.65-1.15)\,p(130)+(8.92-1.65)\,p(140)=0.0021+0.0250=0.0271.$$
This is the adjusted-envelope identity of Theorem~\ref{thm:general-polyhedral-layer-pricing} applied to the layers at $130$ and $140$. The second term is the catastrophe bond's multiplier $\mu_{\mathrm{ILS}}$ at $C_{\mathrm{GOV}}=0$. It makes up $92\%$ of the total, so most of the value of public capacity comes from using the existing catastrophe bond higher in the tower. If we compared the pool and the backstop at $130$, as if the backstop replaced the pool there, we would obtain $(8.92-1.15)\,p(130)=0.032$ and overstate the value of public capacity by about $20\%$.

For $0<C_{\mathrm{GOV}}<35$, the backstop covers $(130,130+C_{\mathrm{GOV}}]$, the catastrophe bond covers $(90,130]$ and $(130+C_{\mathrm{GOV}},140+C_{\mathrm{GOV}}]$, and the pool keeps the losses above $140+C_{\mathrm{GOV}}$; Figure~\ref{fig:cat-tower-allocation} is the case $C_{\mathrm{GOV}}=15$. One more unit of public capacity triggers the same two substitutions, now at $130+C_{\mathrm{GOV}}$ and $140+C_{\mathrm{GOV}}$, so that
$$\mu_{\mathrm{GOV}}(C_{\mathrm{GOV}})=0.50\,p(130+C_{\mathrm{GOV}})+7.27\,p(140+C_{\mathrm{GOV}}),$$
with $p$ evaluated at the lower end of each marginal layer. As $C_{\mathrm{GOV}}$ grows, both substitutions move further into the tail, where exceedance probabilities are smaller, and the demand curve decreases: $\mu_{\mathrm{GOV}}$ is $0.023$ at $C_{\mathrm{GOV}}=10$ and $0.016$ at $C_{\mathrm{GOV}}=30$. Because the layers have width $0.1$, the multiplier changes slightly every time a coverage boundary crosses a layer boundary, which is why the curve looks continuous in the figure.

At $C_{\mathrm{GOV}}=35$, the catastrophe bond ($50$) and the backstop ($35$) fill the whole band $(90,175]$, of width $85$. Above $175$, only the pool and the reinsurers are eligible, and the reinsurers have used their $52$ units below $90$, so the pool keeps $(175,250]$; this is the floor of $75$ in the right panel. More public capacity can still replace the catastrophe bond, but the capacity freed at the catastrophe bond can no longer replace the pool. Only the first substitution remains, and the demand curve drops from $0.0152$ to $0.50\,p(165)=0.0011$. At $C_{\mathrm{GOV}}=45$, the backstop covers its whole band $(130,175]$, and further capacity has no value.

The middle panel of Figure~\ref{fig:cat-tower-incidence} shows the catastrophe bond's multiplier along the same sweep. For $C_{\mathrm{GOV}}<35$, it equals the second term of the demand curve, $\mu_{\mathrm{ILS}}=7.27\,p(140+C_{\mathrm{GOV}})$. It falls from $0.025$ at $C_{\mathrm{GOV}}=0$ to $0.014$ just below $35$, and it is zero beyond $35$, when the catastrophe bond can no longer replace the pool. Public capacity and catastrophe-bond capacity are therefore substitutes: the more public capacity there is, the less an extra unit of catastrophe bond is worth. A larger backstop and a larger catastrophe bond should be compared at the current size of the other.

\subsection{Model uncertainty}\label{subsec:cat-tower-regime}

With several catastrophe models, a treaty that is cheap under one model may be expensive under another. We therefore solve the robust program~\eqref{eq:robust-solver-primal} with the three scenarios of Table~\ref{tab:cat-tower-scenarios}. The planner chooses a single treaty that minimizes the largest of its three total risk assessments. This is the maxmin criterion of \citet{gilboa1989maxmin} applied to minus the total risk assessment. By Corollary~\ref{cor:robust-lp-dual}, a scenario can receive a positive dual weight $\omega_a^\star$ only if the treaty's assessment under that scenario equals the robust value. We call a scenario with positive weight a pricing scenario, because its costs enter the supporting prices through the weighted costs $H_{ki}^{\bm\omega^\star}$.

\begin{figure}[ht]
\centering
\includegraphics[width=\textwidth]{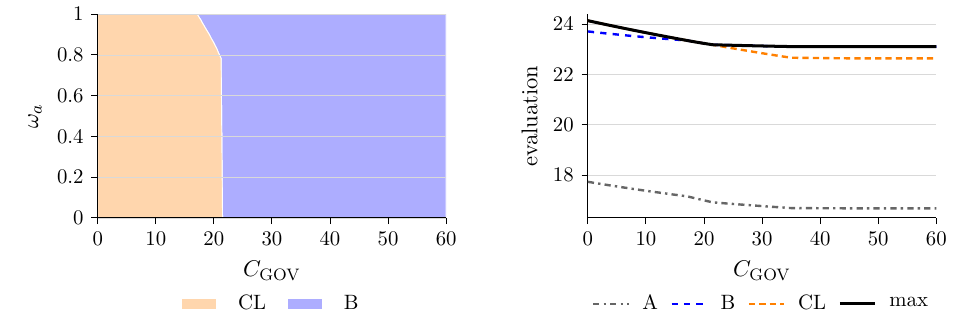}
\caption{Robust program. Left: dual weights of the scenarios $\P^{\mathrm{CL}}$ and $\P^{\mathrm B}$. Right: total risk assessment of the robust treaty under each scenario, and their maximum.}
\label{fig:cat-tower-regime}
\end{figure}

The nominal scenario never receives weight. All participants use the same distortion in every scenario, and $\P^{\mathrm{CL}}$ gives at least as much probability as $\P^{\mathrm A}$ to exceeding every loss level. Any treaty is therefore assessed at least as severely under $\P^{\mathrm{CL}}$ as under $\P^{\mathrm A}$, and strictly more severely at the robust optimum, so $\omega_{\mathrm A}^\star=0$. The relevant comparison is between $\P^{\mathrm B}$ and $\P^{\mathrm{CL}}$. Figure~\ref{fig:cat-tower-regime} shows the weights of these two scenarios (left) and the assessments of the robust treaty under all three scenarios (right).

For $C_{\mathrm{GOV}}\le17$, the climate-conditioned scenario is the worst case and receives all the weight. It puts the most probability on the high layers, where capacity is scarce. Public capacity covers these layers, so as $C_{\mathrm{GOV}}$ increases, the assessment under $\P^{\mathrm{CL}}$ falls faster than the assessment under $\P^{\mathrm B}$, which puts more probability on the lower layers. The two assessments become equal at about $C_{\mathrm{GOV}}=17.1$. Between $17.1$ and $21.5$, both scenarios receive positive weight, and the weight of $\P^{\mathrm B}$ rises from $0$ to about $0.23$ (it is $0.14$ at $C_{\mathrm{GOV}}=20$). Over this range, a change in the treaty lowers the robust value only if it lowers both assessments, so the costs of both scenarios enter the supporting prices. Beyond $21.5$, $\P^{\mathrm B}$ receives all the weight. The scenarios themselves do not change; the switch comes from the change in the robust treaty as public capacity grows.

The robust demand curve in Figure~\ref{fig:cat-tower-demand} follows these weights. At $C_{\mathrm{GOV}}=0$, a unit of public capacity is worth $0.050$ under the robust criterion, $1.86$ times its nominal value of $0.027$. We call the difference the ambiguity premium on the first unit of public capacity. While $\P^{\mathrm{CL}}$ is the worst case, the robust marginal value of public capacity stays close to twice its nominal marginal value, because the backstop covers layers to which $\P^{\mathrm{CL}}$ gives relatively high probability. At $C_{\mathrm{GOV}}=21.5$, the robust capacity multiplier drops from $0.034$ to $0.006$, below the nominal value at the same capacity ($0.019$). Under $\P^{\mathrm B}$, the layers that additional public capacity can take have small exceedance probabilities, so moving them to the backstop saves little. Model uncertainty thus raises the marginal value of public capacity when the backstop is small and lowers it when the backstop is large. As in the nominal case, the robust marginal value of further public capacity is zero beyond $C_{\mathrm{GOV}}=45$.

To choose the size of the backstop, a public authority would compare the demand curve with the marginal cost of providing capacity, expressed on the same basis, and expand the backstop as long as the benefit exceeds the cost. Our linear program provides the benefit, that is, the reduction in the nominal or worst-case total risk assessment after the whole tower is re-optimized. The cost must be specified outside the model. In this example, the benefit declines as the backstop grows, and it drops sharply when the binding constraints or the worst-case scenario change. At each capacity, the dual solution also shows which constraint prevents further risk transfer.

\section{Discussion and conclusion}\label{sec:discussion}

Operational constraints are part of most risk-sharing arrangements used in practice. However, they break the layerwise rule that solves the unconstrained problem under distortion risk measures, so the constrained problem has no closed-form solution in general. We have shown that, for comonotonic allocations on an ordered finite loss grid, the constrained risk-sharing problem is nevertheless a linear program for any distortion functions and any polyhedral system of operational constraints (Theorem~\ref{thm:lp}). The primal solution gives the optimal allocation of the loss layers. The dual solution gives supporting layer prices and a multiplier for each operational constraint, and Theorem~\ref{thm:general-polyhedral-layer-pricing} shows how the multipliers of binding constraints shift the participants' layer costs. These multipliers measure the marginal value of relaxing a constraint after the whole allocation is re-optimized. Because the constrained planner value can have kinks where the set of binding constraints changes, the classical envelope theorem does not apply at those points, and Proposition~\ref{prop:one-sided-sensitivity} gives the one-sided effects of relaxing and tightening a constraint. With finitely many probability and distortion scenarios, the robust problem remains a linear program. The dual weights of the robust linear program identify the scenarios that determine the worst case and hence the robust prices. We use the catastrophe tower to show how these tools measure the value of capacity when several constraints interact.

The allocation problem alone is not a complete economic pricing arrangement, which must also specify admissible transfers or trades and participant budgets. Extensions involving cooperative stability may build on the dual characterizations of core nonemptiness in \citet{bondareva1963applications} and \citet{shapley1967balanced}, as well as the construction of core allocations from optimal prices in linear production games by \citet{owen1975core}, and on the design of entry-monotone risk-sharing pools in \citet{blier-wong2026transfers}.

This paper solves the allocation problem on finite loss grids. If the aggregate loss has a continuous distribution on a bounded range, we can discretize it by rounding each loss to a nearby grid point in a way that preserves the ordering of losses. In Appendix~\ref{sec:discretization}, we show that, as the largest distance between successive grid points tends to zero, the minimum total risk assessment converges to the continuous comonotonic value $V(S;\Lset)$, with an error of at most $n$ times that distance, provided the constraints are preserved when evaluating continuous allocations on the grid and interpolating grid allocations between grid points.

The approach has limitations. It optimizes over comonotonic allocations, which is without loss of optimality when comonotonic improvement preserves the operational constraints. The grid must cover the tail of the aggregate loss, and in the case study the robust results depend on how losses above the grid are treated. The ambiguity set is finite, so the robust value does not account for mixtures of the candidate models. Finally, the model has a single period. Addressing these limitations would make the method more widely applicable, but it would require different techniques, and we leave it for future work.

\section*{Statements and Declarations}

\paragraph{Funding} CB gratefully acknowledges financial support from the Research Foundation Flanders (FWO, G093024N and S006721N). CBW gratefully acknowledges financial support from the Natural Sciences and Engineering Research Council of Canada (RGPIN-2025-06879).

\paragraph{Competing interests} The authors have no competing interests to declare that are relevant to the content of this article.

\paragraph{Availability of data and materials} No datasets were used in this study. All results of the case study in Section~\ref{sec:numerical-examples} are computed from the parameters reported in Tables~\ref{tab:cat-tower-scenarios} and~\ref{tab:cat-tower-agents}.

\paragraph{Code availability} The R code that reproduces the case study in Section~\ref{sec:numerical-examples} is available from the authors upon request.

\appendix
\section{Consistency of ordered-grid approximations}\label{sec:discretization}

In the main text, we fix an ordered finite grid and solve the planner problem on that grid exactly. In this appendix, we ask whether the grid values approximate the planner value $V(S;\Lset)$ of a bounded continuous-loss model as the grid is refined. The argument rests on two Lipschitz facts, both standard. A distortion risk measure is $1$-Lipschitz in the supremum norm, and every admissible contract function is $1$-Lipschitz by Lemma~\ref{lem:contract-function-lipschitz}. Together they show that evaluating a continuous contract on the grid, or interpolating a grid contract linearly between knots, moves the objective by at most $n$ times the mesh. The resulting error bound holds for every distortion function. It covers the unconstrained problem and, once the operational breakpoints are placed on the grid, every admissibility system built from the rows of Table~\ref{tab:constraint-catalogue}. Throughout, $S$ is bounded; truncating an unbounded loss is a separate approximation problem not treated here.

\subsection{Two Lipschitz ingredients}\label{subsec:lipschitz-ingredients}
Recall that $(\Omega,\mathcal{F},\P)$ is an atomless probability space, and $\mathcal{X}:=L^\infty(\Omega,\mathcal{F},\P)$ denotes the space of bounded random variables endowed with the sup-norm.
\begin{lemma}\label{lem:distortion-supnorm}
For every distortion function $h$ and all $Y,Z\in\mathcal X$,
$$|\rho_h(Y)-\rho_h(Z)|\le\|Y-Z\|_\infty.$$
\end{lemma}

\begin{proof}
Every monotone cash-additive functional on $\mathcal X$ has this property; see \citet[Lemma~4.3]{follmer2016stochastic}. With $\varepsilon:=\|Y-Z\|_\infty$, we have $Z-\varepsilon\le Y\le Z+\varepsilon$ almost surely. Monotonicity and cash additivity give $\rho_h(Z)-\varepsilon\le\rho_h(Y)\le\rho_h(Z)+\varepsilon$.
\end{proof}

Combining the no-sabotage requirement with full allocation yields a useful property of the contract functions. In a two-agent insurance model, \citet{carlier2003insurance} give the analogous observation: an indemnity is 1-Lipschitz whenever both the indemnity and retained loss are nondecreasing.
\begin{lemma}\label{lem:contract-function-lipschitz}
If nondecreasing functions $f_1,\dots,f_n:\mathcal D_S\to\R$ satisfy $\sum_{i=1}^nf_i(s)=s$ at every $s\in\mathcal D_S$, then each $f_i$ is 1-Lipschitz and consequently continuous.
\end{lemma}

\begin{proof}
Choose $s,t\in\mathcal D_S$ with $s<t$. Monotonicity of every admissible $f_i$ gives $f_i(t)-f_i(s)\ge0$ for each $i\in[n]$. Adding these nonnegative increments and using clearing yields
$$\sum_{j=1}^n\left(f_j(t)-f_j(s)\right)=t-s.$$
No individual increment can exceed their sum, and hence
$$0\le f_i(t)-f_i(s)\le t-s,\qquad i\in[n].$$
If necessary, exchange $s$ and $t$. We then obtain $|f_i(t)-f_i(s)|\le |t-s|$ for all $s,t\in\mathcal D_S$, proving the claim.
\end{proof}

For an allocation $\bm X\in\mathcal X^n$, write $J(\bm X):=\sum_{i=1}^n\rho_{h_i}(X_i)$. Applying Lemma~\ref{lem:distortion-supnorm} agent by agent gives
\begin{equation}\label{eq:objective-lipschitz}
|J(\bm X)-J(\bm Y)|\le n\max_{i\in[n]}\|X_i-Y_i\|_\infty,
\qquad \bm X,\bm Y\in\mathcal X^n.
\end{equation}
The second ingredient is Lemma~\ref{lem:contract-function-lipschitz}. If $\bm X=(f_1(S),\dots,f_n(S))\in\mathbb A_n^+(S)$, then each $f_i$ is $1$-Lipschitz on $\mathcal D_S$. Hence, for every random variable $S'$ with values in $\mathcal D_S$ we have
\begin{equation}\label{eq:contract-lipschitz}
\|f_i(S')-f_i(S)\|_\infty\le\|S'-S\|_\infty,\qquad i\in[n].
\end{equation}

\subsection{Ordered grids, evaluation, and interpolation}\label{subsec:evaluation-interpolation}

Let $\overline s:=\mathop{\rm ess\,sup}S$ and fix a representative with $0\le S\le\overline s$. For each $m$, choose a knot set
$$\mathcal G_m=\{0=S_0^m<S_1^m<\cdots<S_{K_m}^m=\overline s\}, \qquad \delta_m:=\max_{1\le k\le K_m}\left(S_k^m-S_{k-1}^m\right),$$
with mesh $\delta_m\to0$, and a nondecreasing map $\psi_m:\mathcal D_S\to\mathcal G_m$ with $\psi_m(0)=0$, $\psi_m(\overline s)=\overline s$, and $|\psi_m(s)-s|\le\delta_m$ for all $s$. Rounding up to the next knot is one such map. Set
\begin{equation}\label{eq:ordered-approximation}
S^m:=\psi_m(S),
\qquad
\|S^m-S\|_\infty\le\delta_m.
\end{equation}
The grid $\Pi_m$ is the partition of $\Omega$ generated by $S^m$, with cumulative losses $S_k^m$ and increments $s_k^m:=S_k^m-S_{k-1}^m>0$. The grids need not be nested. Two conventions reconcile this construction with Section~\ref{ss:comonotonic-grid-constraints}. Both are absorbed by defining the layer costs directly from the law of $S^m$:
\begin{equation}\label{eq:appendix-layer-costs}
H_{ki}^m:=h_i\bigl(\P(S^m\ge S_k^m)\bigr),\qquad k=1,\dots,K_m,\quad i\in[n].
\end{equation}
First, when $\P(S^m=0)>0$, the main text lists the zero state as $S_1=0$ with $s_1=0$; that layer forces $x_{1i}=0$ for all $i$, and we suppress it by indexing the knots from $S_0^m=0$. The zero state remains an event of $\Pi_m$ but no longer carries a layer, so the cumulative probabilities start at $P_0^m:=\P(S^m=0)$ rather than at zero, and the convention $P_0=0$ of Section~\ref{ss:comonotonic-grid-constraints} does not apply after this reindexing. In particular, the first positive layer $(0,S_1^m]$ has cost $H_{1i}^m=h_i\bigl(1-\P(S^m=0)\bigr)$, which equals $h_i(1)=1$ only when $\P(S^m=0)=0$. This is the coefficient that Section~\ref{ss:comonotonic-grid-constraints} attaches to the first layer of positive size, so suppressing the zero-size layer leaves the program unchanged. Second, a knot that $S^m$ never attains corresponds to an empty cell. By \eqref{eq:appendix-layer-costs}, such a knot only splits a layer into two layers with equal unit costs $H_{ki}^m=H_{k+1,i}^m$, so the unconstrained value in \eqref{eq:planner-grid} is unchanged. Knots of this kind let $\Lset_{\Pi_m}$ carry an operational breakpoint exactly.

With these costs, the grid objective of a contract is the risk of its payoff. Let $Y=g(S^m)$ with $g$ nondecreasing on $\mathcal G_m$ and $g(0)=0$, and write $Y_k:=g(S_k^m)$ and $y_k:=Y_k-Y_{k-1}$ for $k=1,\dots,K_m$. If $Y_{k-1}<Y_k$ and $t\in[Y_{k-1},Y_k)$, monotonicity of $g$ gives $\{Y>t\}=\{S^m\ge S_k^m\}$. Since $Y\ge0$, the Choquet representation \eqref{eq:choquet} yields
\begin{equation}\label{eq:appendix-grid-linearity}
\rho_h(Y)=\int_0^\infty h\bigl(\P(Y>t)\bigr)\d t=\sum_{k=1}^{K_m}y_k\,h\bigl(\P(S^m\ge S_k^m)\bigr).
\end{equation}
This identity replaces Lemma~\ref{lem:comonotonic-grid-linearity} on $\Pi_m$ and does not require any cell to have positive probability.

\begin{figure}[ht]
\centering
\includegraphics[width=\textwidth]{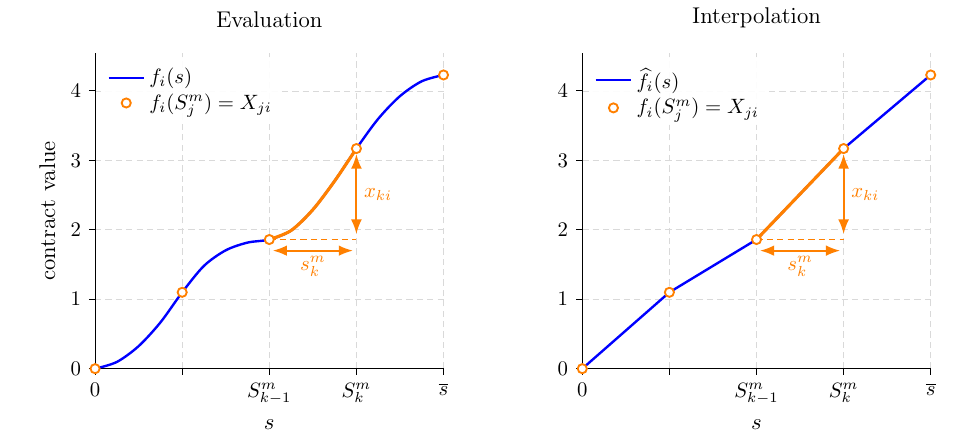}
\caption{Evaluation and interpolation on the aggregate-loss grid.}
\label{fig:evaluation-interpolation}
\end{figure}

A grid contract is a layer vector $\bm x\in\mathcal C_{\Pi_m}(\varnothing)$, with cumulative values $X_{ki}=\sum_{\ell\le k}x_{\ell i}$ and $X_{0i}=0$. Its payoff is $X_i=\sum_kX_{ki}\id_{A_k}$, and by \eqref{eq:appendix-grid-linearity} its objective $\sum_{k=1}^{K_m}\sum_{i=1}^nH_{ki}^mx_{ki}$ in \eqref{eq:planner-grid} equals $J(\bm X)$. Two maps connect continuous and grid contracts (Figure~\ref{fig:evaluation-interpolation}).
\begin{enumerate}[label=(\alph*)]
\item \emph{Evaluation} sends a continuous contract $f=(f_1,\dots,f_n)$ with $f(S)\in\mathbb A_n^+(S)$ to the layer vector $x_{ki}:=f_i(S_k^m)-f_i(S_{k-1}^m)$. Its cumulative values are $f_i(S_k^m)$ and its payoff is $f_i(S^m)$.
\item \emph{Interpolation} sends a grid contract $\bm x$ to the piecewise-affine functions $\widehat f_i$ through the points $(S_k^m,X_{ki})$, $k=0,\dots,K_m$. Its continuous payoff is $\widehat f_i(S)$, and $\widehat f_i(S^m)=X_i$ recovers the grid payoff.
\end{enumerate}

\begin{lemma}\label{lem:evaluation-interpolation}
Evaluation maps $\mathbb A_n^+(S)$ into $\mathcal C_{\Pi_m}(\varnothing)$ with $\|f_i(S^m)-f_i(S)\|_\infty\le\delta_m$ for every $i$. Interpolation maps $\mathcal C_{\Pi_m}(\varnothing)$ into $\mathbb A_n^+(S)$ with $\|\widehat f_i(S)-\widehat f_i(S^m)\|_\infty\le\delta_m$ for every $i$.
\end{lemma}

\begin{proof}
For evaluation, the functions $f_i$ are nondecreasing, vanish at zero, and sum to the identity on $\mathcal D_S$. Hence $x_{ki}\ge0$ and $\sum_{i=1}^n x_{ki}=s_k^m$, so $\bm x\in\mathcal C_{\Pi_m}(\varnothing)$. The bound is \eqref{eq:contract-lipschitz} with $S'=S^m$ and \eqref{eq:ordered-approximation}.

For interpolation, the slope of $\widehat f_i$ on $[S_{k-1}^m,S_k^m]$ is $x_{ki}/s_k^m$, which lies in $[0,1]$ because $x_{ki}\ge0$ and $\sum_{i=1}^n x_{ki}=s_k^m$. Each $\widehat f_i$ is therefore nondecreasing and $1$-Lipschitz with $\widehat f_i(0)=0$. The sum $\sum_{i=1}^n\widehat f_i$ equals the identity at every knot and is affine between knots, so it equals the identity on $\mathcal D_S$. Thus $\widehat f(S)\in\mathbb A_n^+(S)$, and the bound follows from the Lipschitz property and \eqref{eq:ordered-approximation}.
\end{proof}

\subsection{Value consistency}\label{subsec:value-consistency}

Lemma~\ref{lem:evaluation-interpolation} settles the unconstrained problem. For a constrained problem, the same two maps must also preserve the operational rows. This is a property of the pair $(\Lset,\Lset_{\Pi_m})$, not of the polyhedrality of $\Lset_{\Pi_m}$ alone, and we name it.

\begin{definition}\label{def:grid-compatible}
An admissibility system $\Lset$ with grid versions $\Lset_{\Pi_m}$ is \emph{grid-compatible} along $(\Pi_m)_m$ if, for every $m$, evaluation maps $\mathbb A_n^{+,\Lset}(S)$ into $\mathcal C_{\Pi_m}(\Lset)$ and interpolation maps $\mathcal C_{\Pi_m}(\Lset)$ into $\mathbb A_n^{+,\Lset}(S)$.
\end{definition}

The empty system is grid-compatible by Lemma~\ref{lem:evaluation-interpolation}. Both conditions are checked one row at a time, so a system assembled from grid-compatible rows is grid-compatible.

\begin{theorem}\label{thm:grid-consistency}
Let $h_1,\dots,h_n$ be arbitrary distortions and $\Lset$ a grid-compatible admissibility system with $\mathbb A_n^{+,\Lset}(S)\ne\varnothing$. Then, for every $m$,
\begin{equation}\label{eq:grid-consistency-bound}
\left|V_{\Pi_m}(\Lset)-V(S;\Lset)\right|\le n\delta_m,
\end{equation}
and hence $V_{\Pi_m}(\Lset)\to V(S;\Lset)$. For $\Lset=\varnothing$, no assumption beyond \eqref{eq:ordered-approximation} is required.
\end{theorem}

\begin{proof}
The grid feasible set is nonempty because evaluation preserves feasibility. Every feasible payoff satisfies $0\le X_i\le\overline s$, so $0\le J\le n\overline s$ on both feasible sets and both values are finite.

Let $f(S)\in\mathbb A_n^{+,\Lset}(S)$. Its evaluation lies in $\mathcal C_{\Pi_m}(\Lset)$ and has payoff $f(S^m)$, so \eqref{eq:objective-lipschitz} and Lemma~\ref{lem:evaluation-interpolation} give
$$V_{\Pi_m}(\Lset)\le J(f(S^m))\le J(f(S))+n\delta_m.$$
Taking the infimum over $f$ yields $V_{\Pi_m}(\Lset)\le V(S;\Lset)+n\delta_m$.

Conversely, let $\bm x\in\mathcal C_{\Pi_m}(\Lset)$ with interpolant $\widehat f$. Then $\widehat f(S)\in\mathbb A_n^{+,\Lset}(S)$, and $\widehat f(S^m)$ is the payoff of $\bm x$, so
$$V(S;\Lset)\le J(\widehat f(S))\le J(\widehat f(S^m))+n\delta_m.$$
Taking the infimum over $\bm x$ yields $V(S;\Lset)\le V_{\Pi_m}(\Lset)+n\delta_m$.
\end{proof}

It remains to check that the operational rows used in this paper are grid-compatible. Each grid row in Table~\ref{tab:constraint-catalogue} is a continuous condition on the contract functions read at the knots or integrated over a layer. The only requirement is that the loss levels at which a row switches on or off are knots of every grid.

\begin{proposition}\label{prop:catalogue-compatible}
Suppose every band endpoint appearing below belongs to every knot set $\mathcal G_m$. Then each of the following rows is grid-compatible, and so is any admissibility system assembled from them.
\begin{enumerate}[label=(\roman*)]
\item Terminal cap on a set $T\subseteq[n]$: $\sum_{i\in T}f_i(\overline s)\le C_T$, with grid row $\sum_{i\in T}X_{K_mi}\le C_T$.
\item Eligible band $[a,b]$ for agent $i$: $f_i$ is constant on $[0,a]$ and on $[b,\overline s]$, with grid rows $x_{ki}=0$ for every layer outside $[a,b]$. A deductible at $a$ is the case $b=\overline s$, since $f_i(0)=0$.
\item Cumulative corridor for agent $i$: $\underline c\,s\le f_i(s)\le\overline c\,s$ on $\mathcal D_S$, with grid rows $\underline c\,S_k^m\le X_{ki}\le\overline c\,S_k^m$ for all $k$.
\item Layer corridor for agent $i$ on a band $[a,b]$: $\underline c\le f_i'\le\overline c$ almost everywhere on $[a,b]$, with grid rows $\underline c\,s_k^m\le x_{ki}\le\overline c\,s_k^m$ for every layer inside $[a,b]$.
\end{enumerate}
\end{proposition}

\begin{proof}
Since band endpoints are knots, every layer lies entirely inside or entirely outside each band. Every $f_i$ is $1$-Lipschitz by Lemma~\ref{lem:contract-function-lipschitz}, hence absolutely continuous, so $x_{ki}=\int_{S_{k-1}^m}^{S_k^m}f_i'(t)\,\d t$ under evaluation.

Evaluation preserves (i) and (iii) because the grid rows are the continuous conditions at knots. It preserves (ii) because $f_i$ is constant on every layer outside $[a,b]$, and (iv) because the integral of $f_i'$ over a layer inside $[a,b]$ lies between $\underline c\,s_k^m$ and $\overline c\,s_k^m$.

Interpolation reproduces every knot value, which gives (i). On each layer, $\widehat f_i$ is affine with slope $x_{ki}/s_k^m$. Zero slopes on the layers outside $[a,b]$ make $\widehat f_i$ constant on $[0,a]$ and on $[b,\overline s]$, which gives (ii), and slopes in $[\underline c,\overline c]$ on the layers inside $[a,b]$ give (iv). For (iii), the functions $s\mapsto\widehat f_i(s)-\underline c\,s$ and $s\mapsto\overline c\,s-\widehat f_i(s)$ are affine on each layer and nonnegative at both endpoints, hence nonnegative on the layer. Finally, the two maps are checked row by row, so any finite collection of such rows is grid-compatible.
\end{proof}

The case study in Section~\ref{sec:numerical-examples} uses rows of types (i), (ii), and (iv) of Proposition~\ref{prop:catalogue-compatible}, with band endpoints $90$, $130$, and $175$; all three are knots of its grid of width $0.1$. Theorem~\ref{thm:grid-consistency} therefore applies to refinements of that grid, for the aggregate loss censored at $250$. The theorem concerns values only. Interpolated grid optimizers are uniformly bounded and $1$-Lipschitz, so the Arzel\`a--Ascoli theorem yields uniformly convergent subsequences whose limits are feasible, and \eqref{eq:objective-lipschitz} shows that every such limit is optimal for the continuous problem; we do not pursue this here.

\section{Linear-programming facts and proofs}\label{sec:proofs}

\subsection{Linear-programming toolbox}\label{ss:lp-toolbox}

In this section, we collect two standard facts about finite linear programs: duality and sensitivity analysis. Consider the generic finite program
\begin{equation}\label{eq:generic-primal}
\min_{\bm{z}\in\R_+^{\ell}}\ \bm{d}^\top\bm{z}
\qquad\textnormal{s.t.}\qquad
B\bm{z}=\bm{u},\qquad G\bm{z}\le\bm{v},
\end{equation}
having $p$ equality rows and $q$ inequality rows. With unrestricted equality multiplier $\bm{\phi}\in\R^p$ and nonnegative multiplier $\bm{\psi}\in\R_+^q$ on the inequality bounds, its Lagrangian is
\begin{equation}\label{eq:generic-lagrangian}
\mathcal L(\bm{z};\bm{\phi},\bm{\psi})
:=\bm{d}^\top\bm{z}
+\bm{\phi}^\top(\bm{u}-B\bm{z})
+\bm{\psi}^\top(G\bm{z}-\bm{v}).
\end{equation}

\begin{lemma}\label{lem:lp-duality}
Program \eqref{eq:generic-primal} has dual
\begin{equation}\label{eq:generic-dual}
\max_{\bm{\phi}\in\R^{p},\ \bm{\psi}\in\R_+^{q}}\ \bm{u}^\top\bm{\phi}-\bm{v}^\top\bm{\psi}
\qquad\textnormal{s.t.}\qquad
\bm{d}-B^\top\bm{\phi}+G^\top\bm{\psi}\ge\bm{0},
\end{equation}
with the following properties.
\begin{enumerate}[label=(\roman*)]
\item (Strong duality) Whenever \eqref{eq:generic-primal} is feasible and has finite value, the primal and dual both attain that same value.
\item (Complementary slackness) A feasible $\bm{z}$ and feasible $(\bm{\phi},\bm{\psi})$ are jointly optimal exactly when $z_j\bigl(\bm{d}-B^\top\bm{\phi}+G^\top\bm{\psi}\bigr)_j=0$ for every variable $j$ and $\psi_a\bigl((G\bm{z})_a-v_a\bigr)=0$ for every inequality row $a$.
\item (Sensitivity) Let $W(\bm{u},\bm{v})$ denote the optimal value of \eqref{eq:generic-primal}. The function $W$ is convex and polyhedral on its finite-value domain. If $(\bm{\phi}^\star,\bm{\psi}^\star)$ is any optimal dual solution, then
$$\bigl(\bm{\phi}^\star,-\bm{\psi}^\star\bigr)\in\partial W(\bm{u},\bm{v}).$$
At every differentiability point,
$$\nabla_{\bm{u}}W(\bm{u},\bm{v})=\bm{\phi}^\star,\qquad \nabla_{\bm{v}}W(\bm{u},\bm{v})=-\bm{\psi}^\star.$$
\end{enumerate}
\end{lemma}

Lemma~\ref{lem:lp-duality} collects the linear-programming results used in the main text and in the proofs below; see \citet[Sections~4.3 and 5.2--5.3]{bertsimas1997introduction}.

\subsection{Proofs for Section~\ref{sec:main}}

\begin{proof}[Proof of Corollary~\ref{cor:layer_assignment}]
Program \eqref{eq:layer_transport_lp} separates by layer. For a fixed $k$, nonnegativity and $\sum_{i=1}^n x_{ki}=s_k$ define a simplex. The linear objective over that simplex is minimized by assigning the layer to an agent attaining $\min_{i\in[n]}H_{ki}=H_k^\wedge$. Summing these independent minima over all layers produces \eqref{eq:value_closed_form} and proves the result.
\end{proof}

\begin{proof}[Proof of Corollary~\ref{cor:dual_layer_prices}]
Since the primal is feasible and has finite value, Lemma~\ref{lem:lp-duality}(i) gives strong duality. The dual constraints are thus equivalent to $\lambda_k\le \min_{i\in[n]}H_{ki}=H_k^\wedge$ for $k=1,\ldots,m.$
Because $s_k\ge0$, the dual objective is maximized by
$\lambda_k^\star=H_k^\wedge$ for every $k$. If $s_k>0$, an optimal
solution cannot satisfy $\lambda_k<H_k^\wedge$, since increasing
$\lambda_k$ to $H_k^\wedge$ would preserve feasibility and strictly
increase the objective.

Finally, $H_1^\wedge=1$, $H_{m+1}^\wedge=0$, and
$k\mapsto H_k^\wedge$ is nonincreasing. Hence $q_k^\star\ge0$ and
$$\sum_{k=1}^m q_k^\star = H_1^\wedge-H_{m+1}^\wedge = 1.$$
\end{proof}

\begin{proof}[Proof of Proposition~\ref{prop:general-polyhedral-layer-dual}]
Apply Lemma~\ref{lem:lp-duality} with $\bm z=\bm x$, $\bm d=\bm c$, $B=E$, $\bm u=\bm r$, $G=A$, and $\bm v=\bm b$. When $\mathcal C_\Pi(\Lset)$ is nonempty, $P_\Pi(\bm{b},\bm{r}_{\Lset})$ attains a finite value by Theorem~\ref{thm:lp}, so Lemma~\ref{lem:lp-duality}(i) gives strong duality.
\end{proof}

\begin{proof}[Proof of Theorem~\ref{thm:general-polyhedral-layer-pricing}]
Pick a primal--dual optimum for the pair in Proposition~\ref{prop:general-polyhedral-layer-dual}.
Holding optimal $\bm{\mu}^{\Lset}$ and $\bm{\eta}^{\Lset}$ fixed, dual feasibility places $\lambda_k^{\Lset}$ below the stated shadow-adjusted envelope. Since $s_k>0$, $\lambda_k^{\Lset}$ cannot lie strictly below it, as increasing $\lambda_k^{\Lset}$ to the envelope would preserve feasibility and strictly increase the objective. This proves the envelope identity
$$\lambda_k^{\Lset} =\min_{i\in[n]} \left( H_{ki}-(E_{\Lset}^\top\bm{\eta}^{\Lset})_{ki} +(A^\top\bm{\mu}^{\Lset})_{ki} \right).$$
Complementary slackness from Lemma~\ref{lem:lp-duality}(ii) then gives
$$x_{ki}^{\Lset} \left( H_{ki}-\lambda_k^{\Lset} -(E_{\Lset}^\top\bm{\eta}^{\Lset})_{ki} +(A^\top\bm{\mu}^{\Lset})_{ki} \right)=0,$$
and positive $x_{ki}^{\Lset}$ therefore forces its reduced cost to vanish, establishing the second assertion.
\end{proof}

\begin{remark}[Zero-size first layer]\label{rem:zero-size-first-layer}
Since $S_1<\cdots<S_m$, only the first layer can have size zero, which happens when $S_1=0$, that is, when $\P(S=0)>0$. Clearing then forces $x_{1i}=0$ for all $i$, the dual objective has a zero coefficient on $\lambda_1$, and optimality alone does not identify that multiplier. Throughout, we select $\lambda_1$ at the lower envelope of the adjusted costs of layer~$1$, as in Theorem~\ref{thm:general-polyhedral-layer-pricing} or its robust counterpart in Section~\ref{sec:robust-lp}. This choice keeps the dual solution optimal, and the envelope identities, and hence the pricing masses, then hold for every layer.
\end{remark}

\begin{proof}[Proof of Corollary~\ref{cor:operational-shadow-values}]
Complementary slackness, Lemma~\ref{lem:lp-duality}(ii), gives the first claim. Apply Lemma~\ref{lem:lp-duality}(iii) with equality right-hand side $(\bm{s},\bm{r}_{\Lset})$ and inequality right-hand side $\bm{b}$, holding $\bm{s}$ fixed. This gives convexity and the stated subgradient. Increasing $\bm{b}$ enlarges the feasible set, so $V_\Pi$ is nonincreasing in every component of $\bm{b}$. At a differentiability point, the subdifferential is a singleton, which gives the derivative formulas.
\end{proof}

\begin{proof}[Proof of Proposition~\ref{prop:one-sided-sensitivity}]
By Proposition~\ref{prop:general-polyhedral-layer-dual}, $V_\Pi(\bm b)$ is the maximum of the dual objective over the dual feasible set, which does not depend on $\bm b$. Along $\bm b+t\bm d$, the dual objective at a fixed $(\bm\lambda,\bm\eta,\bm\mu)$ is affine in $t$ with slope $-\bm d^\top\bm\mu$. Standard sensitivity analysis therefore gives the right derivative as the largest slope among optimal dual solutions, $\max_{\bm\mu\in\mathcal M(\bm b)}(-\bm d^\top\bm\mu)=-\min_{\bm\mu\in\mathcal M(\bm b)}\bm d^\top\bm\mu$. Applying this result to the direction $-\bm d$ gives the left derivative, and differentiability requires the two one-sided derivatives to agree.
\end{proof}

\subsection{Proofs for Section~\ref{sec:robust-lp}}

\begin{proof}[Proof of Theorem~\ref{thm:robust-lp}]
For a fixed $\bm{x}\in\mathcal C_\Pi(\Lset)$, the smallest $\tau$ that satisfies the scenario rows is
$$\tau=\max_{a\in[M]}\sum_{k=1}^m\sum_{i=1}^n H_{ki}^a x_{ki},$$
so minimizing $\tau$ minimizes the worst-case total risk assessment. All constraints are linear because the scenarios enter through the fixed arrays $(H_{ki}^a)_{k,i}$. If $\mathcal C_\Pi(\Lset)$ is nonempty, it is compact (Section~\ref{ss:comonotonic-grid-constraints}), and the maximum of finitely many linear functions is continuous on it, so the minimum is attained.
\end{proof}

\begin{proof}[Proof of Corollary~\ref{cor:robust-lp-dual}]
Write $\tau=\tau^+-\tau^-$ with $\tau^\pm\ge0$ and apply Lemma~\ref{lem:lp-duality} in the variables $(\bm{x},\tau^+,\tau^-)$. The equality rows have multipliers $(\bm\lambda,\bm\eta)$, the operational inequalities have multipliers $\bm{\mu}\ge\bm{0}$, and the scenario rows have multipliers $\bm{\omega}\ge\bm{0}$. The dual constraints for $\tau^+$ and $\tau^-$ are $1-\sum_{a=1}^M\omega_a\ge0$ and $\sum_{a=1}^M\omega_a-1\ge0$, so $\bm{\omega}\in\Delta_M$. For $x_{ki}$, the scenario rows contribute $\sum_{a=1}^M\omega_aH_{ki}^a=H_{ki}^{\bm{\omega}}$, which gives the stated dual constraint. If $\mathcal C_\Pi(\Lset)$ is nonempty, Theorem~\ref{thm:robust-lp} gives a finite primal optimum, and Lemma~\ref{lem:lp-duality}(i) gives strong duality. For (i), fix $(\bm\eta^\star,\bm\mu^\star,\bm\omega^\star)$: if $\lambda_k^\star$ were below the stated minimum for a layer with $s_k>0$, increasing it would keep the dual feasible and raise the dual objective. Statements (ii) and (iii) are the complementary-slackness conditions of Lemma~\ref{lem:lp-duality}(ii) for $x_{ki}$ and for the scenario rows.
\end{proof}

\begin{proof}[Proof of Corollary~\ref{cor:robust-envelope-rule}]
Without operational rows, the dual~\eqref{eq:robust-solver-dual} maximizes $\bm s^\top\bm\lambda$ subject to $\lambda_k\le H_{ki}^{\bm\omega}$ and $\bm\omega\in\Delta_M$. For fixed $\bm\omega$, the best choice is $\lambda_k=H_k^{\bm\omega,\wedge}$, and strong duality gives~\eqref{eq:robust-unconstrained-envelope-value}. Set $\lambda_k^\star:=H_k^{\bm\omega^\star,\wedge}$ for every $k$. Then $(\bm\lambda^\star,\bm\omega^\star)$ is dual feasible and attains the dual value, so it is dual optimal. If $\bm x$ is optimal, complementary slackness with this dual optimum gives (a) and (b), as in Corollary~\ref{cor:robust-lp-dual}(ii)--(iii). Conversely, if $\bm x$ satisfies (a) and (b), let $\tau$ be its largest total risk assessment across scenarios. Then $(\bm x,\tau)$ and $(\bm\lambda^\star,\bm\omega^\star)$ are feasible and satisfy complementary slackness, so they are optimal by Lemma~\ref{lem:lp-duality}(ii).
\end{proof}

\subsection{Proofs for Section~\ref{sec:caps}}

\begin{proof}[Proof of Proposition~\ref{prop:capped-transport-dual}]
Use Proposition~\ref{prop:general-polyhedral-layer-dual} with one capacity row per agent and no operational equalities. Agent $i$'s row has coefficient one on every $x_{ki}$ and zero on all variables of other agents, with right-hand side $C_i$. Consequently $(A^\top\bm{\mu})_{ki}=\mu_i$, and the general dual becomes \eqref{eq:capped_layer_transport_dual}. Theorem~\ref{thm:general-polyhedral-layer-pricing} provides both the formula for the adjusted envelope and its equality on positive assignments. Complementary slackness for cap $i$ gives $\mu_i^C(\sum_{k=1}^m x_{ki}^C-C_i)=0$, so $\mu_i^C>0$ forces the cap to bind.
\end{proof}

\begin{proof}[Proof of Corollary~\ref{cor:scarcity-adjusted-pricing}]
Proposition~\ref{prop:capped-transport-dual}, strong duality, and complementary slackness yield the envelope, using Remark~\ref{rem:zero-size-first-layer} for a first layer of zero size. Each sequence $k\mapsto H_{ki}$ is nonincreasing, and adding the constant $\mu_i^C$ preserves that ordering. Their pointwise minimum $k\mapsto\lambda_k^C$ is therefore nonincreasing. Since $H_{1i}=h_i(1)=1$ and $\min_i\mu_i^C=0$, we have $\lambda_1^C=1$. Also $\lambda_m^C\ge0$ because both $H_{mi}$ and $\mu_i^C$ are nonnegative. Taking $\lambda_{m+1}^C:=0$, the differences $q_k^C=\lambda_k^C-\lambda_{k+1}^C$ are nonnegative and telescope to one, whence $\Q^C$ is a probability measure.
\end{proof}

\begin{proof}[Proof of Corollary~\ref{cor:single-capacity-value}]
Apply Proposition~\ref{prop:one-sided-sensitivity} to $v_J$ at $\bm C(\kappa)$ with the direction $\bm d=\bm e_i$, which raises agent $i$'s capacity. Since $G(\kappa)=v_J(\bm C(0))-v_J(\bm C(\kappa))$, the one-sided derivatives of $G$ are the negatives of those of $\kappa\mapsto v_J(\bm C(\kappa))$, which gives both formulas. The two one-sided derivatives agree if and only if $\mu_i$ is constant across optimal dual pairs, yielding the differentiable formula.
\end{proof}

\begin{proof}[Proof of Proposition~\ref{prop:one-cap-greedy}]
The output $\bm x$ is feasible: every layer is fully assigned, agent~$n$ receives at most $C$, and each uncapped agent receives at most $S_m\le C_i$. By Lemma~\ref{lem:lp-duality}(ii), it suffices to pair $\bm x$ with a feasible point of the dual \eqref{eq:capped_layer_transport_dual} satisfying complementary slackness.

Lift agent~$n$'s curve by the largest advantage it does not fully take: set
$$\mu_n:=\max\bigl(\{0\}\cup\{\Delta_k:\ x_{kn}<s_k\}\bigr),\qquad \mu_i:=0\ \ (i\in[n-1]),\qquad \lambda_k:=\min\bigl(H_{kn}+\mu_n,\ \overline H_k\bigr).$$
This point is dual feasible, since $\bm\mu\ge\bm0$ and $\lambda_k\le H_{ki}+\mu_i$ for all $k,i$. Complementary slackness has three conditions. First, $x_{kn}>0$ requires $\lambda_k=H_{kn}+\mu_n$, that is, $\Delta_k\ge\mu_n$; this holds because the algorithm fills layers in decreasing order of $\Delta_k$, so every layer agent~$n$ touches has advantage at least that of every layer it leaves unfilled. Second, $x_{ki}>0$ for $i\in[n-1]$ requires $\lambda_k=H_{ki}$; here $H_{ki}=\overline H_k$ and $x_{kn}<s_k$, so $\Delta_k\le\mu_n$ and $\lambda_k=\overline H_k$. Third, $\mu_n>0$ requires the cap to bind; $\mu_n>0$ means a layer with positive advantage is not full, which happens only once capacity has run out.
\end{proof}

\end{document}